\documentclass[b4paper,10pt,twoside]{article}

\usepackage[latin1]{inputenc} 
\usepackage[T1]{fontenc} 
\usepackage{a4wide}
\usepackage{amsmath,amsthm}
\usepackage{amssymb}
\usepackage{a4wide}
\usepackage{graphicx,epsfig}
\usepackage{color}
\usepackage{epic}
\usepackage{enumerate}

\def\CC{\mathbb C}
\newcommand\dps {\displaystyle }
\newcommand{\br}{{\bold r}}
\newcommand{\ri}{{\rm i}}
\newcommand{\rx}{{x,{\rm ELP}}}

\newcommand{\xS}{{x,S}}
\newcommand{\mR}{{{\mathbb{R}}^3}}
\newcommand{\RR}{{{\mathbb{R}}}}

\newcommand{\LL}{L}
\newcommand{\phiHF}{\phi}
\newcommand{\psiHF}{\psi}
\newcommand{\rhoHF}{\rho_\Phi}

\newcommand{\KHF}{K_\Phi}

\newcommand{\PHF}{\gamma_\Phi}

\newcommand{\HS}{{\mathfrak{S}_2}}
\newcommand{\tc}{{\mathfrak{S}_1}}

\def\NN{\mathbb N}
\def\div{{\rm div} \,}
\def\Tr{{\rm Tr} \, }

\def\sqw{\hbox{\rlap{\leavevmode\raise.3ex\hbox{$\sqcap$}}$%
\sqcup$}}
\def\cqfd{\ifmmode\sqw\else{\ifhmode\unskip\fi\nobreak\hfil
\penalty50\hskip1em\null\nobreak\hfil\sqw
\parfillskip=0pt\finalhyphendemerits=0\endgraf}\fi}

\numberwithin{equation}{section}
\newtheorem{theoreme}{Theorem}[section]
\newtheorem{proposition}[theoreme]{Proposition}

\newtheorem{lemme}[theoreme]{Lemma}

\newtheorem{hyp}[theoreme]{Assumption}

\begin{document}

\title{Local Exchange Potentials for Electronic Structure Calculations}
\author{
Eric Canc\`{e}s$^{1,2}$, Gabriel Stoltz$^{1,2}$, Gustavo E. Scuseria$^3$, \\ 
Viktor N. Staroverov$^4$, Ernest R. Davidson$^{5}$ \\
\footnotesize{1: CERMICS, Ecole Nationale des
  Ponts et Chauss\'{e}es (ParisTech), 6 \& 8 Av. Pascal,
  77455 Marne-la-Vall\'{e}e, France} \\
\footnotesize{2: MICMAC project, INRIA, Domaine de Voluceau,
  B.P. 105, 78153 Le Chesnay Cedex, France} \\
\footnotesize{3: Department of Chemistry, Rice University, Houston,
  Texas 77005} \\
\footnotesize{4: Department of Chemistry,  The University of Western
  Ontario, London, Ontario N6A 5B7, Canada}\\
\footnotesize{5: Department of Chemistry, University of Washington,
  Seattle, Washington 98195}\\
}

\maketitle

\begin{abstract}
The Hartree-Fock exchange operator is an integral operator arising in
the Hartree-Fock method and replaced by a multiplicative operator
(a local potential) in Kohn-Sham density functional theory.
This article presents a detailed analysis of the mathematical properties
of various local approximations to the nonlocal Hartree-Fock exchange
operator, including the Slater potential, the optimized effective
potential (OEP), the Krieger-Li-Iafrate (KLI) and
common energy-denominator approximations (CEDA) to the OEP, and the
effective local potential (ELP).  
In particular, we show that the Slater, KLI, CEDA potentials and the ELP 
can all be defined as solutions to certain variational problems. We also
provide a rigorous derivation of the integral OEP equation
and establish the existence of a solution to a system of coupled
nonlinear partial
differential equations defining the Slater approximation to the
Hartree-Fock equations.
\end{abstract}

%
%

\section{Introduction}
The Hartree-Fock exchange operator associated 
with a first-order electron density matrix
$\gamma \in H^1(\RR^3 \times \RR^3)$ is 
a Hilbert-Schmidt operator on $L^2(\RR^3)$ defined by 
\begin{equation}
\forall \phi \in L^2(\RR^3), \quad
(K \phi)(\br) = - \int_{\RR^3} \frac{\gamma(\br,\br')}{|\br-\br'|} \,
\phi(\br') \, d\br'
\end{equation}
(see e.g. \cite{CDKLM03,LS77,Lions87,Solovej} for mathematically
oriented introductions to the Hartree-Fock model).
The construction of this nonlocal (integral) operator is the most
computationally demanding step in solving the Hartree-Fock
equations, especially for periodic systems.
In the early days of quantum chemistry,
Slater proposed to approximate the operator $K$ by a local 
(i.e. multiplicative) operator $v_x(\br)$, termed the exchange
potential~\cite{Slater51}. Nowadays, the nonlocality of the Hartree-Fock 
exchange operator is rarely seen as an obstacle for numerical calculations,
at least in Gaussian basis sets. Recent advances in the development
of orbital-dependent Kohn-Sham density
functionals~\cite{reviewOEP00,WeitaoYang07,Gorling07,ISSDSC07,KummelPerdew03}
have caused a resurgence of interest in representing the exchange interaction
by a local potential. Local exchange potentials
have also been used as input in time-dependent and linear response
methods for computing excitation
energies and other properties~\cite{KG02,VanLeeuwen96,TDDFT,UGG95}. 

The first local approximation to the Hartree-Fock exchange
operator originally proposed by Slater in 1951~\cite{Slater51} is given by
\begin{equation} \label{eq:Slater_potential_first}
v_{x,S}(\br) = - \frac{1}{\rho(\br)} \int_{\RR^3}
\frac{|\gamma(\br,\br')|^2}{|\br-\br'|} \, d\br',
\end{equation}
where $\rho(\br) = \gamma(\br,\br)$ is the electron density.
The second approximation put forth in the same paper is
\begin{equation} \label{eq:Xalpha}
v_{x,{\rm X}\alpha}(\br) = - C_x \rho(\br)^{1/3},
\end{equation}
where $C_x$ is a positive constant.

Motivated by Slater's work, Sharp and Horton~\cite{SH53} introduced
a variational method for obtaining local potentials that approximate
the Hartree-Fock exchange operator. Considering a local potential
$W$ and the associated one-electron Schrödinger operator $H_W =
-\frac12 \Delta + W$, they defined the energy functional $E(W)$
as the Hartree-Fock energy of the Slater determinant constructed
from the lowest $N$~eigenfunctions of $H_W$. The local potential
$W^{\rm OEP}$ which minimizes $E(W)$ is called the optimized
effective potential (OEP).  The exchange part $v_{x,{\rm OEP}}$
of $W^{\rm OEP}$ obtained by subtracting from $W^{\rm OEP}$ the
nuclear and electronic Coulomb potentials can then be used as an
approximation to the Hartree-Fock exchange operator. This idea
was elaborated by Talman and Shadwick~\cite{TS74}.  Unfortunately,
it is difficult to give a proper mathematical formulation of the
(infinite-dimensional) OEP search problem or to build consistent
finite-dimensional approximations to the OEP (of course, the
two issues are closely related). An alternative is to solve the
Euler-Lagrange equation of the OEP 
minimization problem, the so-called integral OEP equation.
Two well-known practical approximations to this equation
are the Krieger-Li-Iafrate (KLI) approximation~\cite{KLI92} and
the common energy-denominator approximation (CEDA)~\cite{GB01}
which yield, respectively, the local exchange potentials
$v_{x,{\rm KLI}}$ and $v_{x,{\rm CEDA}}$.

More recently, several other techniques for generating local exchange
potentials have been proposed~\cite{HC,SSD06}. These techniques
associate a density matrix $\gamma$ with an effective local potential
(ELP) that provides a variational approximation to the corresponding
nonlocal Hartree-Fock exchange operator.  The construction of the
Slater potential or an ELP requires a density matrix $\gamma$,
whereas the OEP, KLI and CEDA potentials are obtained by solving
a self-consistent system of equations in which the density matrix
is unknown.  It has been pointed out~\cite{HC,SSD06} that the CEDA
potential coincides with the potential obtained by iterating the
ELP procedure until the orbitals and the ELP are consistent with
each other. The CEDA potential and the self-consistent ELP also
coincide with the localized Hartree-Fock (LHF) potential introduced
in Ref.~\cite{DsG01} \newline

The purpose of this article is to study the mathematical properties of
the local exchange potentials described above. Let us emphasize
that the mathematical problems under examination are nontrivial and
that only partial results have been rigorously established so far. 
Although several articles containing mathematical statements about
local potentials have been published, very few of those articles
involve mathematically rigorous
arguments, and some contain erroneous claims. Many current beliefs about
local exchange potentials rest on {\it implicit} assumptions
(such as the existence and uniqueness of the solution,
differentiability of a functional, existence of a limit, convergence of
an asymptotic expansion, etc.) that cannot be or have not been
formally proven. Our results may seem weaker than those
presented elsewhere, but at least they are rigorous.
\newline

The article is organized as
follows. In Section~\ref{sec:HF}, the main mathematical properties of
the Hartree-Fock model are briefly reviewed. The Slater
potential~(\ref{eq:Slater_potential_first}) is dealt with in
Section~\ref{sec:Slater}. We provide a variational characterization
of the Slater potential constructed from a given $\gamma$ and study
its asymptotic behavior. Then we focus on
self-consistent equations obtained by replacing
the Hartree-Fock exchange operator with the Slater potential
(\ref{eq:Slater_potential_first}) in 
the Hartree-Fock equations. These equations, first
written by Slater~\cite{Slater51}, do not seem to be the Euler-Lagrange
equations of any optimization problem. Therefore, we use a fixed-point
method to prove that, in the radial case (one nucleus at the origin
and spherically symmetric orbitals), they actually have a 
solution. Note that the situation is completely different
if one uses potential (\ref{eq:Xalpha}) instead
of (\ref{eq:Slater_potential_first}). In that case, the self-consistent
equations have an unambiguous variational interpretation: they are the Kohn-Sham
equations obtained with the exchange-correlation functional
\[
E_{x,{\rm X}\alpha} (\rho) = - \frac{3C_x}4 \int_{\RR^3}
\rho^{4/3}(\br) \, d\br,
\]
which has been extensively studied from a mathematical point of view
in Ref.~\cite{LeBris93}. 

In Section~\ref{sec:OEP}, we focus on the OEP. We summarize the known
mathematical results for the OEP problem and provide
a rigorous derivation of the OEP integral equation. We also study
the KLI and CEDA potentials. In Section~\ref{sec:ELP} we prove that
the self-consistent ELP coincides with the CEDA potential. We do not
provide a complete mathematical study of the self-consistent KLI,
CEDA and ELP equations, but only examine the analytical properties of
the corresponding exchange potentials. We prove that, given a set
of molecular orbitals and under some technical assumptions (always
satisfied in practice), the KLI potential and the ELP (hence the
CEDA potential) are uniquely defined, up to an additive constant.

In order to keep the notation as simple as possible, we focus in
Sections~\ref{sec:HF}-\ref{sec:ELP} on fully spin-polarized
electron densities, i.e., systems with spin-up 
electrons only. With the notable exceptions of one-electron systems and
two-electron triplet states, very few systems of practical
interest are fully spin-polarized. However, the mathematical results stated in
Sections~\ref{sec:HF}-\ref{sec:ELP} are completely general in the sense that they
can be easily adapted to closed-shell and spin-polarized systems. Details
are given in Section~\ref{sec:spin}. 

All the proofs are postponed until Section~\ref{sec:proofs}. 
Basic concepts of functional analysis that are necessary
to understand our arguments are summarized in the Appendix
for the non-mathematical readership.

%
%

\section{Hartree-Fock exchange operator}
\label{sec:HF}

Let us first recall the mathematical formulation
of the Hartree-Fock model. As we deal with fully spin-polarized
systems here, the spin variable can be omitted. Without loss
of generality, we will also assume real-valued functions. In the
Hartree-Fock setting, the electronic state 
of a system of $N$ electrons is described by a collection $\Phi = \left(
  \phi_i \right)_{1 \le i \le N}$ of $N$ $L^2$-orthonormal orbitals:
\begin{equation} \label{eq:ortho_constraints}
\int_{\RR^3} \phi_i(\br) \, \phi_j(\br)  \, d\br = \delta_{ij},
\end{equation}
or, equivalently, by the density matrix 
$$
\gamma_\Phi(\br,\br') = \sum_{i=1}^N \phi_i(\br) \phi_i(\br'),
$$
the electronic density being given by
\begin{equation}
\label{eq:1RDM}
\rho_\Phi(\br) = \gamma_\Phi(\br,\br) = \sum_{i=1}^N |\phi_i(\br)|^2. 
\end{equation}
Denoting by 
\begin{equation}
\label{eq:general_nuclear_potential}
V_{\rm nuc}(\br) = - \sum_{k=1}^K \frac{z_k}{|\br-{\bold R}_k|}
\end{equation}
the potential generated by the nuclei ($z_k$ is the charge of the $k$-th
nucleus, ${\bold R}_k$ its position), the Hartree-Fock functional reads
\begin{eqnarray*}
E^{\rm HF}(\Phi) & = & \dps \frac12 \sum_{i=1}^N \int_\mR |\nabla
\phi_i(\br)|^2 \, d\br + \int_\mR V_{\rm nuc}(\br) \,  \rho_\Phi(\br) \,
d\br + \frac12 \int_\mR \int_\mR
\frac{\rho_\Phi(\br)\rho_\Phi(\br')}{|\br-\br'|} \, d\br \, d\br' 
\\ & & 
- \frac12 \int_\mR \int_\mR \frac{|\gamma_\Phi(\br,\br')|^2}{|\br-\br'|} \,
d\br \, d\br'.
\end{eqnarray*}
Each term of the Hartree-Fock energy functional
is well-defined provided $\Phi \in (H^1(\RR^3))^N$, that is, provided
$\nabla \phi_i \in (L^2(\RR^3))^3$ for all $1 \le i \le N$, or in other
words, provided the kinetic energy of $\Phi$ is finite. This property
results from the inequalities~\cite{D76}
\begin{equation} \label{eq:T_Ene}
\left| \int_{\RR^3} \frac{\rho_\Phi(\br)}{|\br-{\bold R}_k|} \, d\br \right| \le 
N^{1/2} \, \left( \sum_{i=1}^N  \int_\mR |\nabla
\phi_i(\br)|^2 \, d\br\right)^{1/2},
\end{equation}
and
\begin{equation} \label{eq:T_Eee}
\int_\mR \int_\mR \frac{|\gamma_\Phi(\br,\br')|^2}{|\br-\br'|} \,
d\br \, d\br' \le 
\int_\mR \int_\mR \frac{\rho_\Phi(\br)\rho_\Phi(\br')}{|\br-\br'|} \, d\br \, d\br'
\le
N^{3/2} \, \left( \sum_{i=1}^N  \int_\mR |\nabla
\phi_i(\br)|^2 \, d\br \right)^{1/2}.
\end{equation}
The Hartree-Fock ground state energy of the system is obtained by
solving the minimization problem
\begin{equation} \label{eq:HF_pb}
I^{\rm HF} = \inf \left\{ E^{\rm HF}(\Phi), \; \Phi \in {\cal X}_N \right\}
\end{equation}
where
\[
{\cal X}_N = \left\{ \Phi = (\phi_i)_{1\le i \le N} \in (H^1(\RR^3))^N
  \; \left | \; \int_{\RR^3} \phi_i(\br) \,  \phi_j(\br) \, d\br=
    \delta_{ij} \right. \right\}. 
\]
The inequalities (\ref{eq:T_Ene}) and  (\ref{eq:T_Eee}) imply
that the Hartree-Fock functional is always bounded from below on ${\cal
  X}_N$.  The Hartree-Fock ground state energy is therefore well-defined
for any molecular system of arbitrary charge. 
The existence of a Hartree-Fock ground state, that is, of some $\Phi^{\rm
  GS} \in {\cal X}_N$ satisfying
$$
E^{\rm HF}(\Phi^{\rm GS}) = \inf \left\{ E^{\rm HF}(\Phi), \; \Phi \in
  {\cal X}_N \right\} 
$$
has been proved by Lieb and Simon~\cite{LS77} for neutral systems
and positive 
ions. It is also known that 'very negative' atomic ions are not
stable: Denoting by $Z$ the charge of the nucleus,
\begin{itemize}
\item (\ref{eq:HF_pb}) has no minimizer when $N > 2Z+1$~\cite{Lieb84}; 
\item there exists a universal constant $Q \ge 0$ (whose optimal value
  is not known) such that for $N \ge Z+Q$, (\ref{eq:HF_pb}) has no
  minimizer~\cite{Solovej}. 
\end{itemize}  
>From a physical
point of view, the instability of very negative ions results from the
fact that the excess 
electrons escape to infinity. Mathematically, it is due to
a loss of compactness at infinity. These two viewpoints can be linked by the
concentration-compactness theory due to
P.L. Lions~\cite{LionsCC1}. The 
existence of a Hartree-Fock ground state for 'moderately' negative ions
is still an open problem (only the stability of H$^-$ has been
mathematically established). 

\medskip

\noindent
The Euler-Lagrange equations associated with the minimization
problem~(\ref{eq:HF_pb}) read
\begin{equation} \label{eq:HF_EL}
{\cal F}_\Phi \phi_i = \sum_{j=1}^N \lambda_{ij} \phi_j,
\end{equation}
where $\Lambda = (\lambda_{ij})$ is a symmetric $N \times N$ matrix (it
is the
Lagrange multiplier of the matrix
constraint~(\ref{eq:ortho_constraints})), 
and where ${\cal F}_\Phi$ is the Fock
operator 
$$
{\cal F}_\Phi = -\frac12 \Delta + V_{\rm nuc}  + 
\rho_\Phi \star \frac{1}{|\br|}  + K_\Phi. 
$$
In the above expression, $\star$ denotes the convolution product:
$$
(f \star g)(\br) := \int_{\RR^3} f(\br') \, g(\br-\br') \, d\br',
$$
and
$K_\Phi$ is the so-called exact-exchange (or 
Hartree-Fock exchange) operator. It is the integral (nonlocal)
operator defined by
\begin{equation} \label{eq:HFexchange}
(K_\Phi \phi)(\br) = - \int_{\RR^3} \frac{\gamma_\Phi(\br,\br')}{|\br-\br'|} \,
\phi(\br') \, d\br'.
\end{equation}
It is easy to see that $K_\Phi$ is a self-adjoint Hilbert-Schmidt operator on
$L^2(\RR^3)$. Indeed, the kernel
$\frac{\gamma_\Phi(\br,\br')}{|\br-\br'|}$
is a square integrable function on $\RR^3 \times \RR^3$~\cite{LS77}.
\newline

For neutral systems and positive ions, ${\cal F}_\Phi$
(for any $\Phi \in {\cal X}_N$) is a self-adjoint operator on
$L^2(\RR^3)$ with domain $H^2(\RR^3)$, and is bounded from below. Its
essential spectrum is $\sigma_{\rm ess}({\cal F}_\Phi) =
[0,+\infty)$. For positive ions, 
the discrete spectrum of ${\cal F}_\Phi$ consists of an infinite
non-decreasing sequence of negative eigenvalues of finite
multiplicities, which converges to zero~\cite{Lions87}. 
\newline

Any minimizer of (\ref{eq:HF_pb}) satisfies the Euler-Lagrange equations
(\ref{eq:HF_EL}). Using the invariance of the Hartree-Fock problem with
respect to orbital rotation~\cite{R51}, it is possible to diagonalize
the matrix 
$\Lambda = [\lambda_{ij}]$ appearing in (\ref{eq:HF_EL}). More precisely,
if $U$ is an 
orthogonal $N \times N$ matrix (i.e. such that $U^T U= UU^T = I_N$) and
if $\Phi \in {\cal X}_N$, then $\Phi U = (\sum_{j=1}^N U_{ji} \phi_j)_{1
  \le i \le N} \in {\cal X}_N$ and $E^{\rm HF}(\Phi U) = E^{\rm HF}(\Phi)$
(in fact $\gamma_{\Phi U} = \gamma_\Phi$, so that one also has ${\cal
  F}_{\Phi U} = {\cal F}_\Phi$). Let $\Phi$ be a solution of
(\ref{eq:HF_EL}) and $U$ an orthogonal $N
\times N$ matrix which diagonalizes the matrix $\Lambda$, i.e. such that 
$U^T \Lambda U = \mbox{Diag}(\eta_1, \cdots, \eta_N)$. Then
$\Psi = (\psi_i)_{1 \le i \le N} = \Phi U$ is a critical point of
(\ref{eq:HF_EL}), with the same energy as $\Phi$, such that for all~$i$,
$$
{\cal F}_\Phi \psi_i = {\cal F}_\Psi \psi_i = \eta_i \psi_i.
$$
This means that $\Psi$ is a collection of $N$ orthonormal eigenvectors
of the Fock operator. Besides, it can be proved that if $\Phi$
is a Hartree-Fock ground state, then 
\begin{itemize}
\item {\it Aufbau} principle~\cite{LS77}: the $\eta_i$'s are the
  lowest $N$ eigenvalues of ${\cal F}_\Phi$ (including multiplicities),
  i.e. $\eta_i = \epsilon_i$ (up to renumbering of the orbitals);
\item No-unfilled-shell property~\cite{BLLS}: $\epsilon_N < \epsilon_{N+1}$,
  i.e. there is always a gap between the highest occupied molecular
  orbital (HOMO) and the lowest unoccupied molecular orbital (LUMO).  
\end{itemize}
Consequently, solving the system
\begin{equation} \label{eq:HFeq}
\left\{ \begin{array}{l}
{\cal F}_\Phi \phi_i = \epsilon_i \phi_i, \\
\Phi=(\phi_i)_{1 \le i \le N} \in {\cal X}_N, \\
\epsilon_1 \le \epsilon_2 \le \cdots \le \epsilon_N \mbox{ are the
  lowest $N$ eigenvalues of ${\cal F}_\Phi$}, 
\end{array} \right.
\end{equation}
and applying orbital rotations to the so-obtained solutions provides all
the global minimizers of~(\ref{eq:HF_pb}), as well as, possibly, local
minimizers and other kinds of critical points.

Let us finally mention that it is possible to reformulate the Hartree-Fock
problem is terms of density operators. Recall that the density operator
$\Upsilon$ associated with the density matrix $\gamma$ is the
self-adjoint operator defined by
$$
(\Upsilon \phi)(\br) = \int_{\RR^3} \gamma(\br,\br') \, \phi(\br') \, d\br'.
$$
In other words, the density matrix $\gamma$ is the kernel of the
integral operator $\Upsilon$. If $\Phi \in {\cal X}_N$, the density
operator $\Upsilon_\Phi$ associated with the density matrix $\gamma_\Phi$
is the rank-$N$ orthogonal projector 
$$
\Upsilon_\Phi = \sum_{i=1}^N |\phi_i\rangle \, \langle \phi_i|.
$$
The Hartree-Fock
energy functional can be written as a functional of the density
operator: 
$$
{\cal E}^{\rm HF}(\Upsilon) = \dps \Tr \left( - \frac 1 2 \Delta \Upsilon
\right) +  \int_\mR V_{\rm nuc}(\br) \rho_\gamma(\br) \, d\br + \frac12
\int_\mR \int_\mR 
\frac{\rho_\gamma(\br)\rho_\gamma(\br')}{|\br-\br'|} \, d\br \, d\br' 
- \frac12 \int_\mR \int_\mR \frac{|\gamma(\br,\br')|^2}{|\br-\br'|} \,
d\br \, d\br', 
$$
where $\gamma$ is the kernel of $\Upsilon$ and $\rho_\gamma(\br) =
\gamma(\br,\br)$.  If $\gamma$ is regular enough,
$$
\Tr \left( - \frac 1 2 \Delta \Upsilon \right) = -\frac 1 2 \int_{\RR^3}
\left. \Delta_\br \gamma(\br,\br') \right|_{\br'=\br} \, d\br.
$$
The above definition of $\Tr \left( - \frac 1 2 \Delta \Upsilon \right)$
can be extended to any non-negative self-adjoint operator $\Upsilon$,
by noting that
$$
\Tr \left( - \frac 1 2 \Delta \Upsilon \right) = \frac 1 2 \Tr \left(
  |\nabla| \Upsilon |\nabla| \right)
$$
when $\gamma$ is regular, and since the operator $|\nabla| \Upsilon
|\nabla|$ is self-adjoint and non-negative, the right-hand side can
always be given a sense in $\RR_+ \cup \left\{+\infty\right\}$ (it equals
the trace of $|\nabla| \Upsilon |\nabla|$ if this operator is
trace-class, and takes the value $+\infty$ otherwise).

\medskip

The Hartree-Fock ground
state energy and density matrices can be obtained by solving
\begin{equation} \label{eq:minHF_DM}
\inf \left\{ {\cal E}^{\rm HF}(\Upsilon), \; \Upsilon \in {\cal P}_N \right\}
\end{equation}
with
$$
{\cal P}_N = \left\{\Upsilon \in {\cal S}(L^2(\RR^3)) \; | \; \Upsilon^2 =
  \Upsilon, \; \Tr( -\Delta \Upsilon) < \infty, \; \Tr(\Upsilon) = N\right\}.
$$
A remarkable property of the Hartree-Fock functional~\cite{LiebVP} is that the
minimizers of (\ref{eq:minHF_DM}) coincide with those of
\begin{equation} 
  \label{eq:minHF_DM_relax}
  \inf \left\{ {\cal E}^{\rm HF}(\Upsilon), \; \Upsilon \in \widetilde {\cal
    P}_N \right\} 
\end{equation}
where
$$
\widetilde {\cal P}_N = \mbox{Convex hull of } {\cal P}_N =
\left\{\Upsilon \in {\cal S}(L^2(\RR^3)) \; | \; 0 \le \Upsilon \le 1
, \; \Tr( -\Delta \Upsilon) < \infty, \; \Tr(\Upsilon) = N\right\}.
$$
Recall that the notation $0 \le \Upsilon \le 1$ means 
$$
\forall \phi \in L^2(\RR^3), \quad 0 \le \langle \phi | \Upsilon |\phi
\rangle \le \| \phi \|^2_{L^2}.
$$
Note that a generic element of $\widetilde {\cal P}_N$ is of the form
$$
\Upsilon = \sum_{i=1}^{+\infty} n_i |\psi_i\rangle \, \langle \psi_i|,
$$
where $(\psi_i)$ is a Hilbert basis of $L^2(\RR^3)$ with $\psi_i \in
H^1(\RR^3)$, $0 \le n_i \le 1$, and $\sum_{i=1}^{+\infty} n_i = N$. This
property is 
the theoretical foundation of efficient algorithms for solving the
Hartree-Fock problem~\cite{CL00b,CDKLM03}. 

\medskip

In what follows, we will denote respectively by $K_\gamma$ and ${\cal
  F}_\gamma$ the Hartree-Fock exchange
operator and the Fock operator associated with the density matrix $\gamma$:
$$
(K_\gamma \phi)(\br) = - \int_{\RR^3} \frac{\gamma(\br,\br')}{|\br-\br'|} \,
\phi(\br') \, d\br', \qquad 
{\cal F}_\gamma = - \frac 1 2 \Delta + V_{\rm nuc} + \rho_\gamma \star
\frac{1}{|\br|} + K_\gamma.
$$

%
%

\section{Slater exchange potential}
\label{sec:Slater}

The Slater exchange potential associated with some $\Phi \in {\cal X}_N$
has the following expression~\cite{Slater51}:
\begin{equation}
\label{eq:Slater_potential}
v^\Phi_\xS(\br) = - \frac{1}{\rho_\Phi(\br)} \, \int_\mR
\frac{|\gamma_\Phi(\br,\br')|^2}{|\br-\br'|} \, d\br'.
\end{equation}
Obviously, the above definition does not make sense if
$\rho_\Phi(\br)=0$. This is not a problem if $\rho_\Phi > 0$ almost
everywhere, since, in view of Proposition~\ref{prop:quali_Slater} below,
(\ref{eq:Slater_potential}) defines an essentially bounded function on
$\RR^3$. If $\rho_\Phi$ vanishes on a set $\Omega$ of positive measure, the
Slater potential will be set to zero on $\Omega$. There is some
arbitrariness here, but as the density of physical systems is positive
almost everywhere, this is not an issue.
\newline

Note that in the case $N=1$ and $\rho_\Phi > 0$ almost everywhere, the
Slater potential is given by
$$
v^\Phi_\xS(\br) = - \frac{1}{|\phi_1(\br)|^2} \, \int_\mR
\frac{|\phi_1(\br) \phi_1(\br')|^2}{|\br-\br'|} \, d\br' = - \int_{\RR^3} 
\frac{|\phi_1(\br')|^2}{|\br-\br'|} \, d\br',
$$
and therefore cancels out the Coulomb potential (this is a case of exact
self-interaction correction).  

The following Proposition collects the main mathematical properties of
the Slater potential associated with a given $\Phi \in {\cal X}_N$. 
\begin{proposition}
\label{prop:quali_Slater}
Let $\Phi = (\phi_i)_{1 \le i \le N} \in {\cal X}_N$.
\begin{enumerate}[(1)]
\item The Slater potential $v^\Phi_\xS$ is an essentially bounded
  function which satisfies almost everywhere
  $$
  - \int_\mR \frac{\rho_\Phi(\br')}{|\br-\br'|} \, d\br' \le v^\Phi_\xS(\br) \le 0.
  $$
  In particular, $v^\Phi_\xS$ vanishes at infinity.
  \newline
  
  Besides, if $\rho_\Phi > 0$ almost everywhere and if one of the
  conditions below is satisfied: 
  \begin{itemize}
  \item the orbitals $\phi_i$ are radial (i.e. spherically symmetric);
  \item there exists $1 \le p < 3/2 < q \le 2$ such that $|\br|\rho_\Phi
    \in L^p(\RR^3) \cap L^q(\RR^3)$,
  \end{itemize}
the asymptotic behavior of the Slater potential is given by
  \begin{equation} \label{eq:decay_vxS}
    v^\Phi_\xS(\br) = -\frac{1}{|\br|} + {\rm o}\left(\frac{1}{|\br|}\right);
  \end{equation}
\item If $\rho_\Phi > 0$ almost everywhere, the Slater potential
  $v^\Phi_\xS$ is the unique minimizer of the variational problems 
  \begin{equation}
    \label{eq:var_Slater}
    \inf \left\{ I_\Phi^{\rm S}(v), \; v \in L^{3}(\mR) + {\rm
      L}^{\infty}(\mR) \right\}  \qquad \mbox{and} \qquad
    \inf \left\{ J_\Phi^{\rm S}(v), \; v \in L^{3}(\mR) + {\rm
      L}^{\infty}(\mR) \right\},
  \end{equation}
  where 
  $$
  I_\Phi^{\rm S}(v)=   \frac12 \|   (v - \KHF) \Upsilon_\Phi \|^2_\HS,
  \quad
  J_\Phi^{\rm S}(v) = \frac12 \|  v  \Upsilon_\Phi - \KHF\|^2_\HS.
  $$
\end{enumerate}
\end{proposition}
Here and below, $\HS$ denotes the vector space of the Hilbert-Schmidt
operators on $L^2(\RR^3)$ and $\| \cdot \|_\HS$ the Hilbert-Schmidt
norm (see~Appendix). In particular
\begin{eqnarray*}
I_\Phi^{\rm S}(v) & = & \frac 1 2 \int_{\RR^3} \int_{\RR^3} \left|
 v(\br) \gamma_\Phi(\br,\br')  + \int_{\RR^3}
  \frac{\gamma_\Phi(\br,\br'') \gamma_\Phi(\br'',\br')}{|\br-\br''|} \, d\br''
\right|^2 \, d\br \, d\br',
\\
J_\Phi^{\rm S}(v) & = & \frac 1 2 \int_{\RR^3} \int_{\RR^3} \left|
 v(\br) \gamma_\Phi(\br,\br')  +
  \frac{\gamma_\Phi(\br,\br')}{|\br-\br'|}  \right|^2 \, d\br \, d\br'.
\end{eqnarray*}

The condition that there exists $1 \le p < 3/2 < q \le 2$ such that 
$|\br|\rho_\Phi \in L^p(\RR^3) \cap L^q(\RR^3)$ obviously holds true when
$\rho_\Phi$ decays exponentially fast, which is the case, in
particular, when $\Phi$ is a Hartree-Fock~\cite{LS77} or a
Kohn-Sham LDA ground state, or a solution to the self-consistent Slater
equation (\ref{eq:SCF_Slater}) (this is a straightforward application of
the maximum principle since Kohn-Sham LDA and Slater potentials vanish at
infinity).  
\newline

In general, $v^\Phi_\xS$ is not a continuous function. This can be seen
writing $v^\Phi_\xS(\br)$ as
$$
v^\Phi_\xS(\br) = - \sum_{i,j=1}^N \frac{\phi_i(\br) \phi_j(\br)}{\rho_\Phi(\br)}
\, \int_{\RR^3} \frac{\phi_i(\br')\phi_j(\br')}{|\br-\br'|} \, d\br'. 
$$
The functions
$$
\br \mapsto \int_{\RR^3} \frac{\phi_i(\br')\phi_j(\br')}{|\br-\br'|} \, d\br'
$$
are continuous, while the
functions $\frac{\phi_i(\br) \phi_j(\br)}{\rho_\Phi(\br)}$ can be
discontinuous at a given point, either because one of the $\phi_i$ is
discontinuous, or because $\rho_\Phi$ vanishes.
It is worth mentioning two special cases.
When $\Phi=(\phi_i)$ is a Hartree-Fock ground state,
it can be proved by elliptic regularity arguments~(see~\cite{LS77}
for instance) that for all $i = 1,\dots,N$, 
$$
\phi_i \in C^\infty(\RR^3 \setminus \left\{ {\bold R}_k \right\}) \cap C^{0,1}(\RR^3).
$$
It then follows that $v^\Phi_\xS$ is globally Lipschitz in any compact set of
$\RR^3 \setminus \rho_\Phi^{-1}(0)$ and $C^\infty$ in 
$\RR^3 \setminus (\rho_\Phi^{-1}(0) \cup \left\{ {\bold R}_k \right\})$. Stronger
regularity can be obtained if $\Phi=(\phi_i)$ is a Kohn-Sham LDA
ground state, or a solution to the self-consistent Slater
equation (\ref{eq:SCF_Slater}). In this case indeed $\rho_\Phi$ is
positive in $\RR^3$ 
(the ground state of the corresponding mean-field Hamiltonian is positive and
non-degenerate), so that $v^\Phi_\xS$ is globally 
Lipschitz in $\RR^3$ and $C^\infty$ away from the nuclei.
\newline

We have not been able to recognize in the self-consistent Slater equations 
\begin{equation} \label{eq:SCF_Slater}
\left\{ \begin{array}{l}
\dps \left( -\frac12 \Delta + V_{\rm nuc} + \rho_\Phi
\star \frac{1}{|\br|} + v^\Phi_\xS  \right) \phi_i = \epsilon_i
\phi_i, \\
\dps \int_{\RR^3} \phi_i(\br) \, \phi_j(\br) \, d\br = \delta_{ij}, \\
\epsilon_1 \le \cdots \le \epsilon_N \mbox{ are the lowest $N$
  eigenvalues of } \left( -\frac12 \Delta + V_{\rm nuc} + \rho_\Phi
\star \frac{1}{|\br|} + v^\Phi_\xS  \right),
\end{array} \right.
\end{equation}
the Euler-Lagrange equations of some minimization 
problem. It is however possible to prove by a fixed point method that
(\ref{eq:SCF_Slater}) has at least one solution for neutral atoms
and positively charged atomic ions, provided only radial orbitals are considered. 
Recall that a function $\phi$ is said to be
radial if there exists a function~$\varphi$ such that $\phi(\br) = \varphi(|\br|)$.
We will denote by $L^2_r(\mR)$ (resp. $H^1_r(\mR)$) the
set of radial $L^2(\mR)$ (resp. radial $H^1(\mR)$) functions, and set 
$$
{\cal X}^r_N = \left\{ \Phi = (\phi_i)_{1\le i \le N} \in (H_r^1(\RR^3))^N
  \; \left | \; \int_{\RR^3} \phi_i(\br) \phi_j(\br) \, d\br =
    \delta_{ij} \right. \right\}. 
$$

\begin{theoreme}
\label{thm:fixed_point_Slater}
In the case of a single nucleus of charge~$Z \geq N$,
(\ref{eq:SCF_Slater}) has a solution\footnote{In the {\it Aufbau}
  condition ''$\epsilon_1 \le \cdots \le \epsilon_N$ are the lowest $N$
  eigenvalues of $\left(-\frac12 \Delta + V_{\rm nuc} + \rho_\Phi
\star \frac{1}{|\br|} + v^\Phi_\xS  \right)$'', the mean-field Hamiltonian 
is here considered as an operator on $L^2_r(\RR^3)$.} $\Phi =(\phi_i) 
\in {\cal X}^r_N$ with $\epsilon_N > 0$ and the corresponding exchange potential
$v^\Phi_\xS$ is globally Lipschitz in $\RR^3$, $C^\infty$ away from the
nucleus, and satisfies, for all $\eta > 0$,
\[
v^\Phi_\xS(\br) = -\frac{1}{|\br|} + {\rm
  o}\left({\rm e}^{- (2 \, \sqrt{-2\epsilon_N}-\eta)|\br|}\right).
\]
Besides, the minimum of the Hartree-Fock energy over the set of the radial
solutions to (\ref{eq:SCF_Slater}) is attained.
\end{theoreme}

In summary, one can associate with any $\Phi \in {\cal X}_N$ a Slater
potential $v^\Phi_\xS$. Among all the local potentials that can be
constructed in this way, we can select a few of them which might be more
physically relevant than the others:
\begin{enumerate}[(i)]
\item the potential $v^{\Phi^{\rm HF}}_\xS$, where $\Phi^{\rm HF}$ is
  a Hartree-Fock ground state of the system;
\item the potential $v^{\Phi^{\rm KS}}_\xS$, where $\Phi^{\rm KS}$ is
  a Kohn-Sham ground state of the system;
\item the potential $v^{\Phi^{\rm SCF}}_\xS$, where $\Phi^{\rm SCF}$ is
  a solution to the self-consistent equations (\ref{eq:SCF_Slater})
  which minimizes the Hartree-Fock energy (over the set of the
  solutions to (\ref{eq:SCF_Slater})). The existence of such $\Phi^{\rm
    SCF}$ is granted in the radial case for neutral atoms and positive
  atomic ions by the last assertion of
  Theorem~\ref{thm:fixed_point_Slater}.  
\end{enumerate}

\section{Optimized Effective Potential  (OEP)}
\label{sec:OEP}

\subsection{Original formulation of the OEP problem}

As already mentioned in the introduction, the OEP approach consists in
minimizing the energy of the Slater determinant constructed with the
lowest $N$~eigenfunctions of some one-electron Schrödinger operator
$H_W = -\frac12 \Delta + W$, $W$ being a ``local potential''.
Note that in most articles dealing with OEP, the set of admissible local
potentials is not defined. 
Leaving temporarily this issue aside, we denote by ${\cal Y}$ a given
set of local potentials (${\cal Y}$ can be for instance the vector space
${\cal Y} = L^{3/2}(\RR^3) + 
L^\infty(\RR^3)$ arising in the mathematical formulation of the density
functional theory~\cite{LiebDFT}). We introduce the set of
admissible local potentials  
\begin{eqnarray*}
{\cal W} =  \bigg\{ W \in {\cal Y} & \bigg| &  H_W = - \frac
1 2 \Delta + W \mbox{ is a self-adjoint operator on $L^2(\RR^3)$, bounded} \\
& & \mbox{from below, with at least $N$ eigenvalues below its
  essential spectrum} \bigg\},
\end{eqnarray*}
and the OEP minimization set
\begin{equation} \label{eq:def_X1}
  {\cal X}_N^{\rm OEP} = \left\{\Phi = \left\{ \phi_i \right\}_{1\le i
    \le N} \; \left | \; 
  \phi_i \in H^1(\RR^3), \; \mbox{(\ref{eq:cond1_OEP})  and
    (\ref{eq:cond2_OEP}) hold for some $W \in {\cal W}$}
  \right. \right\},
\end{equation}
where conditions (\ref{eq:cond1_OEP}) and (\ref{eq:cond2_OEP}) are defined as
\begin{equation} \label{eq:cond1_OEP}
\left\{ \begin{array}{l}
\dps H_W \phi_i = \epsilon_i \phi_i, \\
\dps \int_{\RR^3} \phi_i(\br)\phi_j(\br) \, d\br = \delta_{ij}, \\
\end{array} \right.
\end{equation}
and
\begin{equation} \label{eq:cond2_OEP}
\epsilon_1 \le \cdots \le \epsilon_N \mbox{ are the lowest $N$
  eigenvalues of } H_W.
\end{equation}
The optimized effective potential problem then reads
\begin{equation} \label{eq:OEP}
I^{\rm OEP} = \inf \left\{ E^{\rm HF}(\Phi), \; \Phi \in {\cal X}_N^{\rm
    OEP} \right\}. 
\end{equation}
Denoting by $\Phi^{\rm OEP}$ a minimizer to (\ref{eq:OEP}), an optimal
effective potential is a local potential $W^{\rm OEP} \in {\cal W}$
which allows one to generate $\Phi^{\rm OEP}$ through
(\ref{eq:cond1_OEP})-(\ref{eq:cond2_OEP}).

In order to emphasize the mathematical issues arising from the
above formulation of the OEP problem, it is worth recalling the general
method for proving the existence of solutions to a minimization problem such
as~(\ref{eq:OEP}). The first step consists in considering a
minimizing sequence, that is, a sequence $(\Phi^n)_{n \in \NN}$
of elements of ${\cal X}_N^{\rm OEP}$ such that
\[
\lim_{n \to \infty} E^{\rm HF}(\Phi^n) = \inf \left\{ E^{\rm HF}(\Phi), \; \Phi \in {\cal X}_N^{\rm OEP} \right\}.
\]
It is easy to check that the sequence $(\Phi^n)_{n \in \NN}$ is bounded
in $(H^1(\RR^3))^N$, hence weakly converges, up to extraction, toward some 
$\Phi^\infty \in (H^1(\RR^3))^N$. It is then standard to prove
(see~\cite{LS77} for instance) that
\begin{equation} \label{eq:SCI}
E^{\rm HF}(\Phi^\infty) \le \inf \left\{ E^{\rm HF}(\Phi), \; \Phi \in
  {\cal X}_N^{\rm OEP} \right\}. 
\end{equation}
The difficult step of the proof is to show that $\Phi^\infty \in {\cal
  X}_N^{\rm OEP}$ (if $\Phi^\infty \in {\cal X}_N^{\rm OEP}$, we can
immediately conclude, using 
(\ref{eq:SCI}), that $\Phi^\infty$ is a solution to~(\ref{eq:OEP})).
For this purpose, we need to introduce a sequence $(W_n)_{n \in \NN}$ of
admissible local potentials ($W_n \in {\cal W}$) such that $\Phi^n$ can be
generated by $W_n$ {\it 
  via} (\ref{eq:cond1_OEP})-(\ref{eq:cond2_OEP}). If the set of local
potentials ${\cal Y}$ is \textit{e.g.} a reflexive Banach space and if $(W_n)_{n
  \in \NN}$ is bounded in ${\cal Y}$, then $(W_n)_{n \in
  \NN}$ converges (up to extraction and in 
some weak sense) to some potential $W_\infty \in {\cal Y}$. We could then try
to pass to the limit in the system
$$
\left\{ \begin{array}{l}
\dps H_{W_n} \phi_i^n = \epsilon_i^n \phi_i^n, \\
\dps \int_{\RR^3} \phi_i^n(\br) \phi_j^n(\br) \, d\br = \delta_{ij}, \\ [10pt]
\epsilon^n_1 \le \cdots \le \epsilon^n_N \mbox{ are the lowest $N$
  eigenvalues of } H_{W_n}, 
\end{array} \right.
$$
using more or less sophisticated functional analysis arguments, in order
to prove that $\Phi^\infty$ satisfies
$$
\left\{ \begin{array}{l}
\dps H_{W_\infty} \phi_i^\infty =
\epsilon_i^\infty \phi_i^\infty,  \\
\dps \int_{\RR^3} \phi_i^\infty(\br) \phi_j^\infty(\br) \, d\br =
\delta_{ij}, \\ [10pt] 
\epsilon^\infty_1 \le \cdots \le \epsilon^\infty_N \mbox{ are the lowest $N$
  eigenvalues of } H_{W_\infty},
\end{array} \right.
$$
hence belongs to ${\cal X}_N^{\rm OEP}$. 
\newline

To make this strategy of proof work, we therefore need to find a
functional space ${\cal Y}$ for which the sequences of local potentials
generating the minimizing sequences
of~(\ref{eq:OEP}) are bounded.
Unfortunately, we have not been able to find any non trivial\footnote{It
is of course possible to construct finite-dimensional functional spaces
${\cal Y}$ for which (\ref{eq:OEP}), with ${\cal X}_N^{\rm OEP}$ defined by
(\ref{eq:def_X1}), has a solution. Reducing artificially the class of
admissible potentials is however not a very satisfactory way to tackle
the OEP problem.} functional space ${\cal Y}$ satisfying the above condition.
This mathematical difficulty has numerical
consequences: It is easy to construct dramatic modifications of the (computed)
  optimized effective potential that are ``almost solutions'' of the OEP
  problem (make the potential oscillate and/or go to infinity at
  infinity), see {\it e.g.}~\cite{SSD06}.

\subsection{A well-posed reformulation of the OEP problem}

A way to circumvent the issues raised in the previous Section is to
replace (\ref{eq:cond1_OEP})-(\ref{eq:cond2_OEP}) with formally
equivalent conditions that do not explicitly refer to a local
potential $W$~\cite{BCL04}. 

Let us first deal with (\ref{eq:cond1_OEP}). For an operator $W$ being
considered as a local potential, it needs to be such that
$$
(W\phi)\psi = (W\psi)\phi
$$
for any regular functions $\phi$ and $\psi$. This
requirement is in fact, at least formally, a characterization of
multiplication operators. Applied to $H_W$, this
characterization reads
\begin{equation} \label{eq:charac_HW}
(H_W \phi)\psi - (H_W\psi)\phi = \frac 1 2 \left( \phi \Delta \psi -
  \psi \Delta \phi \right) =  \frac 1 2 \div(\phi \nabla \psi - \psi
\nabla \phi). 
\end{equation}
It is then clear that if $\Phi =
(\phi_i ) \in (H^1(\RR^3))^N$ satisfies
(\ref{eq:cond1_OEP}) with an operator $H_W$ for which
(\ref{eq:charac_HW}) holds true, we also have
\begin{equation} \label{eq:cond1_OEP_eq}
\left\{ \begin{array}{l}
\dps \div (\phi_i \nabla \phi_1 -  \phi_1 \nabla \phi_i) = c_i \phi_1 \phi_i, \\
\dps \int_{\RR^3} \phi_i(\br)\phi_j(\br) \, d\br  = \delta_{ij}, \\
\end{array} \right.
\end{equation}
with $c_i= 2(\epsilon_i-\epsilon_1)$. Conversely, if $\Phi =
\left\{ \phi_i \right\} \in (H^1(\RR^3))^N$ satisfies
(\ref{eq:cond1_OEP_eq}), then, {\em at least formally}, $\Phi$ satisfies
(\ref{eq:cond1_OEP}) with for instance 
\begin{equation} \label{eq:reconstruct_W}
W = \frac{\dps \sum_{i=1}^N \phi_i \Delta \phi_i + \sum_{i=2}^N
  c_i\phi_i^2}{2 \, \rho_\Phi}, 
\end{equation}
$\epsilon_1=0$, and $\epsilon_i=c_i/2$ for $2 \le i \le N$. 

The idea then is to replace condition (\ref{eq:cond1_OEP}) with the
formally equivalent condition (\ref{eq:cond1_OEP_eq}), which does not
explicitly refer to any local potential. 

In order to account for condition (\ref{eq:cond2_OEP}) in the same way,
we remark that 
for any $\Phi \in {\cal X}_N$ and all $1 \le i \le N$,
$$
\forall \psi \in C^\infty_0(\RR^3), \quad \frac 1 2 \int_{\RR^3}
\phi_i(\br)^2\,|\nabla\psi(\br)|^2 \, d\br =  \langle  \psi\phi_i | \left( 
H_W-\epsilon_i \right) | \psi\phi_i \rangle,
$$
where $C^\infty_0(\RR^3)$ is the set of compactly supported
$C^\infty(\RR^3)$ functions.
It follows from the above equality (see~\cite{BCL04} for details) that
conditions (\ref{eq:cond1_OEP})-(\ref{eq:cond2_OEP}) are
equivalent to  
\begin{equation*}
\left\{ \begin{array}{lll}
\dps H_W \phi_i = \epsilon_i \phi_i, \\
\dps \int_{\RR^3} \phi_i(\br)\phi_j(\br) \, d\br= \delta_{ij}, \\
\forall \psi \in C^\infty_0(\RR^3), \; \forall 1 \le i \le N-1, \\
\dps \int_{\RR^3}\phi_i(\br)^2\,|\nabla\psi(\br)|^2 \, & 
\!\!\!\!\!\!\!\!\!\!\!\!\!\!\!\!\!\!\!\!\!\!\!\!\!\!\!\!\!\!\!\! \ge &
\!\!\!\!\!\!\!\!\!\!\!\!\!\!\!\!\!\!\!\!\!\!\!\!\!\!\!\!\!\!\!\! 
\dps 2 \sum_{j=1}^i (\epsilon_j-\epsilon_1) \left( \int_{\RR^3} \psi(\br)
  \phi_i(\br) \phi_j(\br) \, d\br \right)^2,  \\ & & 
\!\!\!\!\!\!\!\!\!\!\!\!\!\!\!\!\!\!\!\!\!\!\!\!\!\!\!\!\!\!\!\!
\dps + 2 (\epsilon_{i+1}-\epsilon_{1})
 \left(\int_{\RR^3}\psi(\br)^2\,\phi_i(\br)^2 \, d\br - \sum_{j=1}^i \left( 
    \int_{\RR^3} \psi(\br) \phi_i(\br) \phi_j(\br) \, d\br \right)^2 \right).
\end{array} \right.
\end{equation*}
Combining the above result with the formal equivalence between 
(\ref{eq:cond1_OEP}) and (\ref{eq:cond1_OEP_eq}) with $c_i =
2(\epsilon_i-\epsilon_1)$, it is natural to introduce the
optimization problem
\begin{equation} \label{eq:OEP_reform}
\widetilde I^{\rm OEP} = \inf \left\{ E^{\rm HF}(\Phi), \; \Phi \in
  \widetilde{\cal X}_N^{\rm OEP} \right\} .
\end{equation}
where 
\begin{eqnarray*}
\widetilde{\cal X}_N^{\rm OEP} & = & \bigg\{ 
\Phi = \left\{ \phi_i \right\}_{1\le i \le N} \; \left | \; 
  \phi_i \in H^1(\RR^3), \; \dps \int_{\RR^3} \phi_i \phi_j =
  \delta_{ij}, \; \exists \, 0 = c_1 \le c_2 \le \cdots \le c_N <
  \infty, \right. \\
& &  \qquad  \forall 2 \le i \le N, \; \div ( \phi_i \nabla \phi_1 -  \phi_1
\nabla \phi_i) = c_i \phi_1 \phi_i, \; \forall 1 \le i \le N-1, \;
\forall \psi \in C^\infty_0(\RR^3),  \\
& & \qquad  \int_{\RR^3}\phi_i^2\,|\nabla\psi|^2 \ge 
\dps \sum_{j=2}^i c_j \left( \int_{\RR^3} \psi
  \phi_i \phi_j \right)^2 
\dps + c_{i+1}
 \left(\int_{\RR^3}\psi^2\,\phi_i^2 - \sum_{j=1}^i \left( 
    \int_{\RR^3} \psi \phi_i \phi_j \right)^2 \right)
\bigg\} .
\end{eqnarray*}
We thus have eliminated any explicit reference to a local
potential. Note that for any reasonable definition of ${\cal Y}$, it holds 
\begin{equation} \label{eq:HF_OEP_1}
{\cal X}^{\rm OEP} \subset \widetilde{\cal X}^{\rm OEP}_N \subset {\cal
  X}_N,
\end{equation}
The connection between the original OEP problem~(\ref{eq:OEP}) and its
reformulation (\ref{eq:OEP_reform}) can therefore be stated as follows:
If $\widetilde \Phi^{\rm OEP} = ( \widetilde \phi_i^{\rm 
      OEP} )_{1 \le i \le N}$ is solution to
  (\ref{eq:OEP_reform}), and if the reconstructed potential
\begin{equation} \label{eq:OEPr}
W^{\rm OEP} = \frac{\dps \sum_{i=1}^N \widetilde \phi_i^{\rm OEP} \Delta
  \widetilde \phi_i^{\rm OEP} + \sum_{i=2}^N c_i|\widetilde
  \phi_i^{\rm OEP}|^2}{2 \, \rho_{\widetilde \Phi^{\rm OEP}}} 
\end{equation}
defines an element of ${\cal W}$,
then $\widetilde \Phi^{\rm OEP}$ is solution to (\ref{eq:OEP}) and
$W^{\rm OEP}$ is an optimized effective potential.

It is proved in~\cite{BCL04} that for a neutral or positively charged
two-electron system, problem~(\ref{eq:OEP_reform}) has at
least one global minimizer $\widetilde \Phi^{\rm OEP}$. Unfortunately,
we have not been able to establish whether or not the reconstructed
potential (\ref{eq:OEPr}) is a well-defined function. 

Let us conclude this section by remarking that (\ref{eq:HF_OEP_1})
yields 
$$
I^{\rm HF} \le \widetilde I^{\rm OEP} \le I^{\rm OEP}. 
$$
A natural question is whether these inequalities are strict or large. We are
only aware of two partial answers to this question:
\begin{itemize}
\item it is proved in~\cite{BCL04} that in the case of a single nucleus
  of charge $Z \ge 2$ and $N=2$ electrons occupying radial orbitals, 
$$
I^{\rm HF} < \widetilde I^{\rm OEP} ;
$$
\item a formal perturbation argument is used in~\cite{IL03} to show that
  $I^{\rm HF} 
  = I^{\rm OEP}$ for non-interacting electrons and that 
  $I^{\rm HF} < I^{\rm OEP}$ in the presence of an infinitesimal
  Coulomb repulsion term. 
\end{itemize}


\subsection{The OEP integral equation and its approximations}

The functional $W \mapsto E^{\rm HF}(\phi_1^W, \cdots , \phi_N^W)$, where
$(\phi_1^W, \cdots , \phi_N^W)$ satisfy 
\begin{equation} \label{eq:VP}
\left\{ \begin{array}{l}
\dps H_{W} \phi_i^W =
\epsilon_i^W \phi_i^W,  \\
\dps \int_{\RR^3} \phi_i^W(\br) \phi_j^W(\br) \, d\br  = \delta_{ij}, \\ [10pt]
\epsilon^W_1 \le \cdots \le \epsilon^W_N \mbox{ are the lowest $N$
  eigenvalues of } H_{W},
\end{array} \right.
\end{equation}
is not
well-defined for two reasons: First, the set of admissible local
potentials has not been properly characterized, and second, (\ref{eq:VP}) may
have several solutions 
if there is no gap between the HOMO and the LUMO of $H_W$. It is
therefore {\it a fortiori} not possible to define the derivative of this
functional. One can however give a rigorous meaning to the functional
and its derivative for local potentials $W$ satisfying the following
assumption. 

\begin{hyp}
\label{hyp:gap} The potential $W$ belongs to $L^2(\RR^3) + L^\infty(\RR^3)$,
and the hamiltonian $H_W$, defined on the domain $D(H_W) = H^2(\RR^3)$, is a
self-adjoint operator on $L^2(\RR^3)$, bounded from below, with at 
least $N$ eigenvalues (including multiplicities) below its essential
spectrum, and there is a gap  
\begin{equation} \label{eq:gap}
\eta = \epsilon_{N+1}^W - \epsilon_{N}^W > 0
\end{equation}
between $\epsilon_N^W$ (the $N$-th eigenvalue of $H_W$) and
$\epsilon_{N+1}^W$ (the $(N+1)$-th eigenvalue of $H_W$, or the bottom of
the essential spectrum if $H_W$ has only $N$ eigenvalues below its
essential spectrum). 
\end{hyp}

Under Assumption~\ref{hyp:gap}, the ground
state density operator of $H_W$ is uniquely defined and satisfies
\begin{eqnarray*}
\Upsilon_W & = & \mbox{arginf} \left\{ \Tr(H_W\Upsilon), \; \Upsilon \in {\cal
    P}_N \right\} \\ & = & \chi_{(-\infty,\epsilon_F]}(H_W),
\end{eqnarray*}
where $\chi_{(-\infty,\epsilon_F]}$ is the characteristic function of
the range $(-\infty,\epsilon_F]$ and
\begin{equation} \label{eq:epsilonF}
\epsilon_F=(\epsilon_{N}^W+\epsilon_{N+1}^W)/2.
\end{equation}
In addition, if $w \in
L^1(\RR^3) \cap L^\infty(\RR^3)$ is such that $\| w \|_{L^\infty} <
\eta/2$, then $W+w$ still is a local potential for which
Assumption~\ref{hyp:gap} holds true (this follows from the Courant-Fischer
formula~\cite{ReedSimon}). In this case,
\begin{eqnarray*}
\Upsilon_{W+w} & = & \mbox{arginf} \left\{ \Tr(H_{W+w}\Upsilon), \;
  \Upsilon \in {\cal 
    P}_N \right\} \\ & = & \chi_{(-\infty,\epsilon_F]}(H_{W+w}),
\end{eqnarray*}
with $\epsilon_F$ given by
(\ref{eq:epsilonF}). It is therefore possible to define the functional
$$
w \mapsto {\cal E}^{\rm HF}(\Upsilon_{W+w})
$$
on the ball
$$
{\cal B}_{\eta/2} = \left\{ w \in L^1(\RR^3) \cap L^\infty(\RR^3), \;  
\| w \|_{L^1 \cap L^\infty} < \eta/2 \right\}.
$$
For $W$ satisfying Assumption~1, one can also define the exchange part of
the potential $W$ as
$$
v_{x}^{W} = W - V_{\rm nuc} - \rho_{\gamma_W} \star \frac{1}{|\br|}
$$
where $\gamma_W$ is the kernel of $\Upsilon_W$.
It is easy to see that $v_{x}^{W} \in L^2(\RR^3) + L^\infty(\RR^3)$.
We are now in position to state the main result of this section.

\begin{theoreme}
\label{thm:integral_OEP} Let $W$ be a local potential such that
Assumption~\ref{hyp:gap} holds true. Then, there exists a unique
function $\varrho^W \in L^1(\RR^3) \cap 
H^2(\RR^3)$ such that
\begin{equation}
\label{eq:full_integral_OEP}
{\cal E}^{\rm HF}(\Upsilon_{W+w}) = {\cal E}^{\rm HF}(\Upsilon_{W}) + \int_{\RR^3} 
\varrho^W(\br) \, w(\br) \, d\br
+ {\rm O}\left (\|w\|^2_{L^1 \cap L^\infty}\right).
\end{equation}
In particular, the function $w \mapsto {\cal E}^{\rm
  HF}(\Upsilon_{W+w})$ is Fr\'{e}chet differentiable at $w=0$. 
Denoting by $R^0(z) = (z-H_W)^{-1}$ the resolvent of $H_W$, by ${\cal
  C}$ a regular closed contour enclosing the lowest $N$ eigenvalues
of $H_W$ (see Figure~1), and by $t_W(\br,\br')$ the kernel of the finite
rank operator
\begin{equation} \label{eq:op_A}
T_W =  \frac{1}{2\pi \ri} \oint_{\cal C} R^0(z) (K_{\gamma_W} - v_x^W) 
R^0(z) \, dz,
\end{equation}
it holds $\varrho^W(\br) = t_W(\br,\br)$. Let $(\phi_i^W)_{1 \le i \le N}$
be a set of $N$ orthonormal eigenvectors of $H_W$ associated with the
lowest $N$ eigenvalues $\epsilon_1^W \le \cdots \le \epsilon_N^W$ of
$H_W$. Then 
$$
\varrho^W(\br) = 2 \, \sum_{i=1}^N \phi_i^W(\br) \; \left[ (1-\Upsilon_W)
  [\epsilon_i^W-(1-\Upsilon_W)H_W(1-\Upsilon_W)]^{-1}
 (1-\Upsilon_W)  (K_{\gamma_W} - v_x^W) \phi_i^W
\right] \!\! (\br).
$$
\end{theoreme}
\begin{figure}[h]
\label{fig:contour}
\centering
\psfig{figure=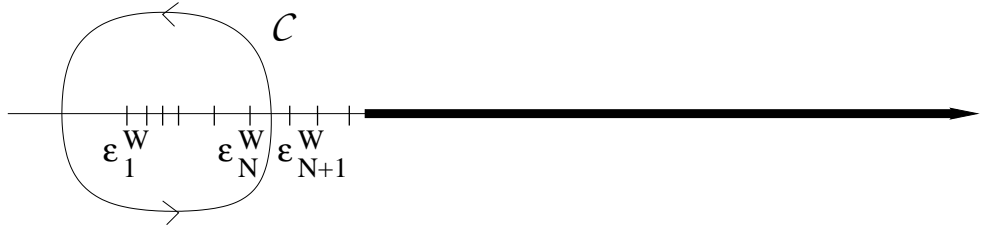,height=3truecm}
\caption{Integration contour ${\cal C}$ in the complex plane}
\end{figure}

Let us come back to the formulation (\ref{eq:OEP}) of the OEP
problem. If no artificial restriction on the set of admissible local
potentials is enforced, ${\cal Y}$ must be such that for all $W \in {\cal
  Y}$, and all $w \in C^\infty_0(\RR^3)$, one also has $W +w \in {\cal Y}$. Let us now
consider a local potential $W \in {\cal Y}$ satisfying
Assumption~\ref{hyp:gap}. Then $W \in {\cal W}$ and it follows from 
(\ref{eq:full_integral_OEP}) that if $W$ is an optimized effective
potential, then
\begin{equation} \label{eq:IOEP}
\varrho^W = 0 .
\end{equation}
Although not obvious at first sight, (\ref{eq:IOEP}) is a
rigorous formulation of the OEP integral equation introduced
in~\cite{SH53,TS74}. 
To clarify this point, we now assume
that the spectrum of $H_W$ is purely discrete (this assumption
is implicit made in~\cite{SH53,TS74}, but is obviously violated
for isolated molecular systems, since for such systems, $W$ is expected
to have a constant, finite value at infinity). In this
case, there exists a Hilbert basis $(\phi_n^W)_{n \in \mathbb{N}}$ of $L^2(\RR^3)$
consisting of eigenvectors of $H_W$ associated with the eigenvalues
$\epsilon_1^W \le \epsilon_2^W \le \cdots \le \epsilon_N^W < \epsilon_{N+1}^W
\le  \cdots$, and the resolvent can be rewritten as
\[
R^0(z) = \sum_{n=1}^{+\infty} \frac{|\phi_n^W \rangle \, \langle
  \phi_n^W|}{z-\epsilon_n^W}.
\]
Then,
\begin{eqnarray*}
&& \left[ (1-\Upsilon_W)
  [\epsilon_i^W-(1-\Upsilon_W)H_W(1-\Upsilon_W)]^{-1}
 (1-\Upsilon_W)  (K_{\gamma_W} - v_x^W) \phi_i^W
\right] \!\! (\br) \\ && = \sum_{a=N+1}^{+\infty} 
\frac{\langle \phi_a^W \, | \,
  K_{\Phi^W}-v_{x}^{W}
  \, | \, \phi_i^W \rangle }{\epsilon_i^W - \epsilon_a^W} \,
\phi_a^W(\br), 
\end{eqnarray*}
so that condition~(\ref{eq:IOEP}) also reads
\begin{equation}
\label{eq:OEP_chemistry}
\sum_{i=1}^N \sum_{a=N+1}^{+\infty} \frac{\langle \phi_a^W \, | \,
  K_{\Phi^W}-v_{x}^{W}
  \, | \, \phi_i^W \rangle }{\epsilon_i^W - \epsilon_a^W} \, \phi_i^W(\br) \,
\phi_a^W(\br) = 0.
\end{equation}
We have found no mathematical result on the existence and
uniqueness of the solutions to the self-consistent OEP integral 
equation in the literature. 
\newline

The exact OEP integral equation is of
little interest in itself, but can be useful to build approximations of
optimized effective potentials. It has indeed be proposed~\cite{SH53,TS74},
instead of approximating the optimization problem~(\ref{eq:OEP}), to
approximate the first order optimality condition~(\ref{eq:OEP_chemistry}). This is the
foundations of the KLI (Krieger-Li-Iafrate) and CEDA (common energy
denominator) approximations.

Starting from the idea of Sharp and Horton~\cite{SH53},  
Krieger, Li and Iafrate proposed the following approximation
of the OEP integral equation:
\begin{equation} \label{eq:KLI_eq}
\sum_{i=1}^N \sum_{j \in \NN^\ast, \, j \neq i} \langle \phi_j^W \, | \,
  K_{\Phi^W}-v_{x,{\rm KLI}}^{\Phi^W} \, | \, \phi_i^W \rangle  \, \phi_i^W(\br) \,
\phi_j^W(\br) = 0.
\end{equation}
We refer to~\cite{KLI92} for details on the derivation of
equation~(\ref{eq:KLI_eq}). Note that the inner sum runs over all
(occupied and virtual) states different from $i$, which implies in
particular that equation (\ref{eq:KLI_eq}) is not invariant with respect
to orbital rotations. In order to cover the general case when $H_W$ has
a non-empty continuous spectrum, it is preferable to rewrite
equation~(\ref{eq:KLI_eq}) as
\begin{equation} 
  \label{q:KLI_eq2}
  \rho_{\Phi^W}(\br)  v_{x,{\rm KLI}}^{\Phi^W}(\br) = - \int_{\RR^3}
  \frac{|\gamma_{\Phi^W}(\br,\br')|^2}{|\br-\br'|} \, d\br' + \sum_{i=1}^{N} 
  \langle \phi_i^W \, | \, v_{x,{\rm KLI}}^{\Phi^W}-K_{\Phi^W} \, | \, \phi_i^W \rangle  \, |\phi_i^W(\br)|^2 .
\end{equation}
The KLI exchange potential $v_{x,{\rm KLI}}^{\Phi^W}$ can then be obtained by solving the
self-consistent equations
\begin{equation} 
  \label{eq:KLI_sys}
  \!\left\{ \begin{array}{l}
    \dps H_W \phi_i^W = \epsilon_i^W \phi_i^W,  \\ 
    \dps \int_{\RR^3} \phi_i^W(\br) \phi_j^W(\br) \, d\br = \delta_{ij},
    \\ [10pt] 
    \epsilon^W_1 \le \cdots \le \epsilon^W_N \mbox{ are the lowest $N$
      eigenvalues of } H_{W}, \\ 
    \dps W = V_{\rm nuc} + \rho_{\Phi^W} \star \frac{1}{|\br|} + v_{x,{\rm KLI}}^{\Phi^W},  \\
    \dps \rho_{\Phi^W}(\br)  v_{x,{\rm KLI}}^{\Phi^W}(\br) = - \int_{\RR^3}
    \frac{|\gamma_{\Phi^W}(\br,\br')|^2}{|\br-\br'|} \, d\br' + \sum_{i=1}^{N} 
    \langle \phi_i^W \, | \, v_{x,{\rm KLI}}^{\Phi^W} - K_{\Phi^W} \, | \, \phi_i^W
    \rangle  \, |\phi_i^W(\br)|^2 \quad {\rm a.e.}
  \end{array} \right.
\end{equation}
Note that if $(\Phi^W,v_{x,{\rm KLI}}^{\Phi^W})$ is solution to the above system,
so is  $(\Phi^W,v_{x,{\rm KLI}}^{\Phi^W}+\lambda)$ for any real constant~$\lambda$.
We have not been able to prove an existence result for
(\ref{eq:KLI_sys}). Assuming that (\ref{eq:KLI_sys})
has a solution $(\Phi^W,v_{x,{\rm KLI}}^{\Phi^W})$ with $\Phi^W \in {\cal
  X}_N$, it is however possible to prove the following:

\begin{proposition}
\label{prop:KLI}
Let  $(\Phi^W,v_{x,{\rm KLI}}^{\Phi^W})$ be a solution to  (\ref{eq:KLI_sys}) such that 
$\Phi^W \in {\cal X}_N$ and $\epsilon_1^W < \min \, \sigma_{\rm
  ess}(H_W)$. Then $\rho_\Phi$ is a continuous, positive 
function on~$\RR^3$, and $v_{x,{\rm KLI}}^{\Phi^W}$ is a continuous, bounded
function on~$\RR^3$. Besides,  
\begin{enumerate}[(1)]
\item the potential $v_{x,{\rm KLI}}^{\Phi^W}$ is the unique solution, up to
  an additive constant, to the minimization problem 
  \begin{equation} 
    \label{eq:var_KLI}
    \inf \left\{ J_{\Phi^W}^{\rm KLI}(v), \; v \in L^{3}(\RR^3) +
    L^\infty(\RR^3) \right\}
  \end{equation}
  where 
  $$
  J_\Phi^{\rm KLI}(v) = \frac 1 2 \left( \left\|  (v-K_\Phi) \Upsilon_\Phi
  \right\|^2_\HS - 
  \sum_{i=1}^N \left|\langle \phi_i|(v-K_\Phi)|\phi_i \rangle\right|^2
  \right).
  $$
  In particular, $\Phi^W$ being given, the KLI potential is uniquely
  defined up to an additive constant;
\item it holds  
  \begin{equation} \label{eq:KLI3}
  v_{x,{\rm KLI}}^{\Phi^W}(\br) = v_{x,S}^{\Phi^W}(\br) + \sum_{i=1}^{N} \left( \alpha^{\Phi^W}_i
  - \langle \phi_i^W \, | K_{\Phi^W} \, | \, \phi_i^W \rangle \right) \,
  \frac{|\phi_i^W(\br)|^2}{\rho_{\Phi^W}(\br)}, 
  \end{equation}
  where $v_{x,S}^{\Phi^W}$ is the Slater potential associated with
  $\Phi^W$ and where $\alpha^{\Phi^W} = (\alpha_i^{\Phi^W}) \in \RR^N$ satisfies
  \begin{equation} 
    \label{eq:linear_system_KLI}
    (I_N - S^{\Phi^W}) \alpha^{\Phi^W} = \beta^{\Phi^W},
  \end{equation}
  with 
  $$
  S^{\Phi^W}_{ij} = \int_{\RR^3} \frac{|{\phi_i^W}(\br)|^2 \,
    |{\phi_j^W}(\br)|^2}{\rho_{\Phi^W}(\br)} \, d\br, \qquad
  \beta_i^{\Phi^W} = \int_{\RR^3}  v_{x,S}^{\Phi^W}(\br)
  |\phi_i^W(\br)|^2 \, d\br - \sum_{j=1}^N
  S^{\Phi^W}_{ij} \langle \phi_j^W \, | K_{\Phi^W} \, | \, \phi_j^W \rangle;
  $$
\item the solutions of the linear system (\ref{eq:linear_system_KLI})
  form a one-dimensional affine space of the form
  $$
  \alpha^{\Phi^W} + \RR \left(1, \cdots , 1 \right)^T.
  $$
  Replacing $\alpha^{\Phi^W}$ with $\alpha^{\Phi^W} + \lambda \left(1, \cdots , 1
  \right)^T$ in~(\ref{eq:KLI3}), amounts to replacing $v_{x,{\rm KLI}}^{\Phi^W}$
  with $v_{x,{\rm KLI}}^{\Phi^W}+\lambda$.
\end{enumerate}
\end{proposition}

Note that
contrarily to the situation encountered with the Slater potential
(see problem~(\ref{eq:var_Slater})), the
quadratic functional $J_\Phi^{\rm KLI}$ is convex (it is non-negative), but not
strictly convex. A consequence of that is the non-uniqueness of
$v_x^{\rm KLI}$, which is only defined up to an additive constant.
\newline

Let us now turn to the CEDA potential introduced by Gritsenko and
Baerends~\cite{GB01}. This approximation consists in replacing
in~(\ref{eq:OEP_chemistry}) the denominators $\epsilon_i^W - \epsilon_a^W$ with a
constant, yielding   
$$
\dps \sum_{i=1}^N \sum_{a=N+1}^{+\infty} \langle \phi_i^W \, | \,
  K_{\Phi^W}-v_{x,{\rm CEDA}}^{\Phi^W} \, | \, \phi_a^W \rangle  \, \phi_i^W(\br) \,
\phi_a^W(\br) = 0.
$$
Here also, it is possible to provide a more explicit formulation of
this equation, still valid when $H_W$ has a non-empty continuous
spectrum:
\begin{equation} 
  \label{eq:CEDA_eq}
  \rho_{\Phi^W}(\br)  v_{x,{\rm CEDA}}^{\Phi^W}(\br) = - \int_{\RR^3}
  \frac{|\gamma_{\Phi^W}(\br,\br')|^2}{|\br-\br'|} \, d\br' + \sum_{i,j=1}^{N} 
  \langle \phi_i^W \, | \, v_{x,{\rm CEDA}}^{\Phi^W} -K_{\Phi^W}\, | \, \phi_j^W
  \rangle  \, \phi_i^W(\br) \, \phi_j^W(\br).
\end{equation}
Let us also mention that the common denominator approximation
amounts to replacing in (\ref{eq:op_A}) the resolvent 
$R^0(z) = (z-H_W)^{-1}$ with the resolvent $R^0_{\rm CEDA}(z) = (z-H_W^{\rm
  CEDA})^{-1}$ of the operator
$$
H_W^{\rm CEDA} = \underline{\epsilon} \Upsilon_{\Phi^W} + \overline{\epsilon}
\left ( 1 - \Upsilon_{\Phi^W} \right ),
$$
where $\underline{\epsilon}$ and $\overline{\epsilon}$ lay respectively
inside and outside ${\cal C}$.

The potential $v_x^{\rm CEDA}$ solves the self-consistent equations
\begin{equation*} \label{eq:CEDA_sys}
\!\left\{ \begin{array}{l}
\dps H_W \phi_i^W =
\epsilon_i^W \phi_i^W,  \\ 
\dps \int_{\RR^3} \phi_i^W(\br) \phi_j^W(\br) \, d\br  = \delta_{ij}, \\ [10pt]
\epsilon^W_1 \le \cdots \le \epsilon^W_N \mbox{ are the lowest $N$
  eigenvalues of } H_{W}, \\ 
\dps W = V_{\rm nuc} + \rho_{\Phi^W} \star \frac{1}{|\br|} + v_{x,{\rm CEDA}}^{\Phi^W},  \\
\dps \rho_{\Phi^W}(\br) v_{x,{\rm CEDA}}^{\Phi^W}(\br) = - \int_{\RR^3}
\frac{|\gamma_{\Phi^W}(\br,\br')|^2}{|\br-\br'|} \, d\br' + \sum_{i,j=1}^{N} 
\langle \phi_i^W \, | \, v_{x,{\rm CEDA}}^{\Phi^W} - K_{\Phi^W} \, | \, \phi_j^W
\rangle  \, \phi_i^W(\br) \, \phi_j^W(\br) \quad {\rm a.e.}
\end{array} \right.
\end{equation*}
To our knowledge, the question of existence and uniqueness of the solution
to the above system is still open.
\newline

It turns out that $v_{x,{\rm CEDA}}^{\Phi^W}$ coincides with the
so-called local Hartree-Fock (LHF) exchange potential $v_x^{\rm LHF}$,
obtained by Della Salla 
and Görling on the basis of completely different arguments
(see~\cite{DsG01} for details). We will see in the next section that it
also equals the self-consistent effective local potential
$v_{x,{\rm ELP}}^{\Phi^W}$~\cite{ISSDSC07}.

%
%

\section{Effective Local Potential (ELP)}
\label{sec:ELP}

The effective local potential associated with a given $\Phi \in {\cal
  X}_N$ was originally defined as the local potential minimizing the
function~\cite{ISSDSC07}
$$
v \mapsto 
S_\Phi(v) = \sum_{i=1}^N \sum_{a=N+1}^{+\infty} \left| \langle \phiHF_i |
(v-\KHF) | \phiHF_a \rangle \right|^2,
$$
$(\phiHF_a)_{a \geq N+1}$ being a Hilbert basis of the orthogonal
of the vector space generated by $(\phiHF_i)_{1\leq i \leq N}$. 
A simple calculation shows that $S_\Phi(v) = J_\Phi^{\rm ELP}(v)$ where
\[
J_\Phi^{\rm ELP}(v) = \frac12 \| [v-\KHF,\Upsilon_\Phi] \|^2_\HS ,
\]
$[A, B]=AB-BA$ denoting the commutator of the operators $A$ and $B$.
An intrinsic formulation of the ELP problem therefore reads
\begin{equation}
  \label{eq:ELP_pb}
  \inf \ \{ J_\Phi^{\rm ELP}(v), \; v  \in \LL^3(\mR) + \LL^\infty(\mR) \}.
\end{equation}

\begin{proposition}
\label{prop:ELP}
Let $\Phi = (\phi_i)_{1 \le i \le N} \in {\cal X}_N$. Any solution
$v_{x,{\rm ELP}}^\Phi$ to (\ref{eq:ELP_pb}) satisfies
\begin{equation} 
  \label{eq:ELP_eq}
  \rho_{\Phi}(\br)  v_{x,{\rm ELP}}^\Phi(\br) = - \int_{\RR^3}
  \frac{|\gamma_{\Phi}(\br,\br')|^2}{|\br-\br'|} \, d\br' + \sum_{i,j=1}^{N} \,
  \left( \langle \phi_i|v_{x,{\rm ELP}}^\Phi|\phi_j\rangle  - \langle
  \phi_i| K_\Phi|\phi_j\rangle \right)
  \phi_i(\br) \, \phi_j(\br) 
\end{equation}
and the symmetric matrix $M^\Phi=[\langle \phi_i|v_{x,{\rm
    ELP}}^\Phi|\phi_j\rangle]$ is solution to the linear system
\begin{equation} 
  \label{eq:ELP_mat}
  (I-A^\Phi) M^\Phi = G^\Phi,
\end{equation}
with
$$
A^\Phi_{kl,ij} = \int_\mR \frac{\phiHF_i(\br) \, \phiHF_j(\br) \,
  \phiHF_k(\br) \, \phiHF_l(\br)}{\rhoHF(\br)} \, d\br ,
\qquad
G^\Phi_{kl} = \int_\mR v_{x,S}^\Phi(\br) \phi_k(\br) \phi_l(\br) \, d\br 
- \sum_{i,j=1}^N A^\Phi_{kl,ij} \langle
\phi_i| K_\Phi|\phi_j\rangle .
$$
Besides, if the orbitals $\phi_i$ are continuous and if the open set
$\RR^3 \setminus \rho_\Phi^{-1}(0)$ is connected, 
then the solutions to (\ref{eq:ELP_mat}) form a one-dimensional affine
set of the form 
$$
M^\Phi + \RR I_N,
$$
so that $v_{x,{\rm ELP}}^\Phi$ is uniquely defined, up to an additive
constant, on the set where $\rho_\Phi > 0$, and can be given arbitrary
values on the set where $\rho_\Phi=0$.
\end{proposition}

Comparing (\ref{eq:ELP_eq}) and (\ref{eq:CEDA_eq}), one immediately recognizes
that the self-consistent CEDA potential and the self-consistent ELP
potential defined by
\begin{equation*} \label{eq:ELP_sys}
\!\left\{ \begin{array}{l}
\dps H_W \phi_i^W =
\epsilon_i^W \phi_i^W,  \\ 
\dps \int_{\RR^3} \phi_i^W(\br) \phi_j^W(\br) \, d\br = \delta_{ij}, \\ [10pt]
\epsilon^W_1 \le \cdots \le \epsilon^W_N \mbox{ are the lowest $N$
  eigenvalues of } H_{W}, \\ 
\dps W = V_{\rm nuc} + \rho_{\Phi^W} \star \frac{1}{|\br|} +  v_{x,{\rm ELP}}^{\Phi^W},  \\
\dps \rho_{\Phi^W}(\br)  v_{x,{\rm ELP}}^{\Phi^W}(\br) = - \int_{\RR^3}
\frac{|\gamma_{\Phi^W}(\br,\br')|^2}{|\br-\br'|} \, d\br' + \sum_{i,j=1}^{N} 
\langle \phi_i^W \, | \,  v_{x,{\rm ELP}}^{\Phi^W} - K_{\Phi^W} \, | \, \phi_j^W
\rangle  \, \phi_i^W(\br) \, \phi_j^W(\br) \quad {\rm a.e.}
\end{array} \right.
\end{equation*}
coincide. As already mentioned, we are not aware of a proof of existence
of the solution to this system. We can however use
Proposition~\ref{prop:ELP} to show that if (\ref{eq:ELP_sys}) has a
solution $(\Phi^W, v_{x,{\rm ELP}}^{\Phi^W})$ with $\Phi^W \in {\cal X}_N$ and if
$\epsilon_1 < \mbox{min} \, \sigma_{\rm ess}(H_W)$, then $ v_{x,{\rm
    ELP}}^{\Phi^W}$ can be obtained from $\Phi^W$ by solving an optimization
problem, which has a unique solution, up to an additive constant (the
proof follows the same lines as the proof of
Proposition~\ref{prop:KLI}: $\phi_1^W$ then is a continuous, positive
function on $\RR^3$, which implies that $\rho_{\Phi^W}$ is positive and that
the above connectivity condition is satisfied).  

\section{Extensions to the General, Unrestricted and Restricted
  Hartree-Fock models}
\label{sec:spin}

In the general Hartree-Fock (GHF) model, each molecular
spin-orbital $\phi_i$ is a complex-valued function\footnote{GHF models
  are of particular interest for systems subjected to magnetic fields;
  for such systems, complex-valued wavefunctions are needed.}  with
spin-up and spin-down components, i.e. $\phi_i \in L^2(\RR^3,\CC^2)$. The
orthonormality constraint (\ref{eq:ortho_constraints}) is replaced with
$$
\int_{\RR^3} \phi_i(\br) \cdot \phi_j(\br)^\ast \, d\br = \delta_{ij},
$$
where
$$
\phi_i(\br) = \left( \begin{array}{c} \phi_i(\br,\uparrow) \\
\phi_i(\br,\downarrow) \end{array} \right) \qquad \mbox{and} \qquad
\phi_i(\br) \cdot \phi_j(\br)^\ast = \sum_{\sigma \in \left\{
    \uparrow,\downarrow \right\}} \phi_i(\br,\sigma) \,
\phi_j(\br,\sigma)^\ast. 
$$ 
The density-matrix can then be represented by a $2 \times 2$ hermitian
matrix
$$
\gamma_\Phi(\br,\br') = \left( \begin{array}{cc}
\gamma_\Phi^{\uparrow\uparrow}(\br,\br') &
\gamma_\Phi^{\uparrow\downarrow}(\br,\br') \\
\gamma_\Phi^{\downarrow\uparrow}(\br,\br') &
\gamma_\Phi^{\downarrow\downarrow}(\br,\br') \\
\end{array} \right)
$$
with
$$
\gamma_\Phi^{\sigma\sigma'}(\br,\br') = \sum_{i=1}^N 
\phi_i(\br,\sigma) \, \phi_i(\br',\sigma')^\ast,
$$
and the electronic density $\rho_\Phi$ is the sum of its spin-up and
spin-down components:
$$
\rho_\Phi(\br) = \rho_\Phi^\uparrow(\br)+\rho_\Phi^\downarrow(\br), \qquad
\rho_\Phi^\uparrow(\br) = \gamma_\Phi^{\uparrow\uparrow}(\br,\br), \qquad
\rho_\Phi^\downarrow(\br) = \gamma_\Phi^{\downarrow\downarrow}(\br,\br).
$$
The Hartree-Fock exchange operator associated with $\gamma_\Phi$ is the
integral operator on $L^2(\RR^3,\CC^2)$ defined by
$$
\forall \phi \in L^2(\RR^3,\CC^2), \quad
(K_\gamma\phi)(\br) = - \int_{\RR^3} \frac{1}{|\br-\br'|}
\gamma(\br,\br') \cdot \phi(\br') \, d\br',
$$
where $\cdot$ denotes the usual matrix-vector product, and local
exchange potentials are $2 \times 2$ hermitian matrices of the form
$$
v(\br) =  \left( \begin{array}{cc}
v^{\uparrow\uparrow}(\br) &
v^{\uparrow\downarrow}(\br) \\
v^{\downarrow\uparrow}(\br) &
v^{\downarrow\downarrow}(\br) \\
\end{array} \right).
$$
The variational definition of the Slater potential given in
Section~\ref{sec:Slater} provides a natural way to define a Slater
potential for the GHF framework: It is the local
potential $v$ which minimizes the Hilbert-Schmidt norm of the operator 
$(v-K_\Phi)\Upsilon_\Phi$. A simple calculation leads to
\begin{eqnarray} \label{eq:SlaterGHF}
v^\Phi_{x,S}(\br) & = &  - R_\Phi(\br)^{-1} \Xi_\Phi(\br) -
\frac{1}{\rho_\Phi(\br)} 
\left[ \Xi_\Phi(\br) - R_\Phi(\br)^{-1} \Xi_\Phi(\br) R_\Phi(\br) \right] \\
& = &  - \Xi_\Phi(\br) R_\Phi(\br)^{-1}  - \frac{1}{\rho_\Phi(\br)}
\left[ \Xi_\Phi(\br) - R_\Phi(\br) \Xi_\Phi(\br) R_\Phi(\br)^{-1}
\right] \nonumber 
\end{eqnarray}
where
$$
R_\Phi(\br) = \gamma_\Phi(\br,\br) \qquad \mbox{and} \qquad
\Xi_\Phi(\br) = \int_{\RR^3} \frac{1}{|\br-\br'|} \gamma_\Phi(\br,\br')
\cdot \gamma(\br,\br')^\ast \, d\br'.
$$

\medskip

Within the unrestricted Hartree-Fock (UHF) model, each molecular spin-orbital
is (generally) chosen real-valued and either spin-up, i.e. $\phi_i(\br) = \left(
  \begin{array}{c} \phi_i^\uparrow (\br) \\ 0 \end{array} \right)$,  or
spin-down, i.e $\phi_i(\br) = \left(
  \begin{array}{c} 0 \\ \phi_i^\downarrow (\br) \end{array} \right)$.
 Denoting by $N_\alpha$ (resp. $N_\beta$)
the number of spin-up (resp. spin-down) orbitals, and ordering the
spin-orbitals in such a way that the first $N_\alpha$ of them are
spin-up, the UHF density matrix reads
$$
\gamma_\Phi(\br,\br') = \left( \begin{array}{cc}
\gamma_\Phi^{\uparrow\uparrow}(\br,\br') &
0 \\
0 &
\gamma_\Phi^{\downarrow\downarrow}(\br,\br') \\
\end{array} \right)
$$ 
with
$$
\gamma_\Phi^{\uparrow\uparrow}(\br,\br') = \sum_{i=1}^{N_\alpha}
\phi_i^\uparrow(\br) \phi_i^\uparrow(\br), \qquad  
\gamma_\Phi^{\downarrow\downarrow}(\br,\br') = \sum_{i=1}^{N_\beta}
\phi_{N_\alpha+i}^\downarrow(\br) \phi_{N_\alpha+i}^\downarrow(\br).
$$
Likewise, the UHF exchange operator is diagonal:
$$
K_\gamma =  \left( \begin{array}{cc}
K_\gamma^{\uparrow\uparrow} &
0 \\
0 & 
K_\gamma^{\downarrow\downarrow}
\end{array} \right) \qquad \mbox{with} \qquad
\forall \phi \in L^2(\RR^3), \quad
(K_\gamma^{\sigma\sigma} \phi)(\br) = 
- \int_{\RR^3} \frac{|\gamma^{\sigma,\sigma}(\br,\br')|^2}{|\br-\br'|}
\phi(\br') \, d\br' .
$$
It is then easy to check that in the UHF setting, the general formula
(\ref{eq:SlaterGHF}) reduces to 
$$
v^\Phi_{x,S}(\br) =  \left( \begin{array}{cc}
v_{x,S}^{\Phi,\uparrow\uparrow}(\br) &
0 \\
0 &
v_{x,S}^{\Phi\downarrow\downarrow}(\br) \\
\end{array} \right)
$$ 
with
$$
v_{x,S}^{\Phi,\sigma\sigma}(\br) = - \frac{1}{\rho^\sigma_\Phi(\br)}
\int_{\RR^3}
\frac{|\gamma_\Phi^{\sigma\sigma}(\br,\br')|^2}{|\br-\br'|} \, d\br',
\qquad
\rho^\sigma_\Phi(\br) =  \gamma_\Phi^{\sigma\sigma}(\br,\br).
$$
One recovers in this way the spin-up and spin-down local potentials
originally introduced by Slater in~\cite{Slater51}.

In closed-shell models, each molecular orbital $\phi_i \in L^2(\RR^3)$
is occupied by one
spin-up and one spin-down electrons. Denoting by $N_p = N/2$ the number
of electron pairs, it holds
$$
\gamma_\Phi^{\uparrow\uparrow}(\br,\br') =
\gamma_\Phi^{\downarrow\downarrow}(\br,\br') = \sum_{i=1}^{N_p} 
\phi_i(\br) \phi_i(\br), \qquad 
\rho^\uparrow_\Phi(\br) =  \rho^\downarrow_\Phi(\br), \qquad
v_{x,S}^{\Phi,\uparrow\uparrow}(\br) = v_{x,S}^{\Phi,\downarrow\downarrow}(\br), 
$$
$$
\gamma_\Phi^{\uparrow\downarrow}(\br,\br') =
\gamma_\Phi^{\downarrow\uparrow}(\br,\br') = 0, \qquad
v_{x,S}^{\Phi,\uparrow\downarrow}(\br) =
v_{x,S}^{\Phi,\downarrow\uparrow}(\br) = 0. 
$$
Proposition~\ref{prop:quali_Slater}, Theorem~\ref{thm:integral_OEP},
Proposition~\ref{prop:KLI}, and Proposition~\ref{prop:ELP} apply {\it
  mutatis mutandis} to the RHF setting, as well as to the spin-up and
spin-down components of the UHF exchange operator and local
potentials. As outlined above from the Slater potential, the variational
characterizations (\ref{eq:var_KLI}) and (\ref{eq:ELP_pb}) of the KLI
and ELP potentials can be used to defined KLI and ELP potentials in the
GHF setting.

%
%

\section{Proofs of the main results}
\label{sec:proofs}

Throughout this section, we denote by $B_R$ the open ball of $\RR^3$ of
radius $R$ 
centered at $0$, i.e. $B_R = \left\{ \br \in \RR^3, \; |\br|< R\right\}$ and
by $B_R^c = \RR^3 \setminus B_R$.

\medskip

In order to simplify the notation, we adopt in this section the usual
abuse of notation consisting in denoting by the same letter an integral
operator and its kernel.

\subsection{Proof of Proposition~\ref{prop:quali_Slater}: Properties of
  the Slater potential}
\label{sec:proofs_Slater}

\noindent
If follows from the Cauchy-Schwarz inequality that
$$
|\gamma_\Phi(\br,\br')|^2 = \left| \sum_{i=1}^N \phi_i(\br) \phi_i(\br') \right|^2
\le \left( \sum_{i=1}^N |\phi_i(\br)|^2 \right) \,
\left( \sum_{i=1}^N |\phi_i(\br')|^2 \right) = \rho_\Phi(\br) \, \rho_\Phi(\br'). 
$$
In the set where $\rho_\Phi > 0$, one therefore has
$$
- \int_{\RR^3} \frac{\rho_\Phi(\br')}{|\br-\br'|} \, d\br' = 
- \frac{1}{\rho_{\Phi}(\br)}  \int_{\RR^3} \frac{\rho_\Phi(\br)
    \rho_\Phi(\br')}{|\br-\br'|} \, d\br' \le - \frac{1}{\rho_{\Phi}(\br)}  
\int_{\RR^3} \frac{|\gamma_\Phi(\br,\br')|^2}{|\br-\br'|} \, d\br' =
v_{x,S}^\Phi(\br) \le 0.
$$
In order to establish the decay property, we rewrite $v_{x,S}^\Phi$ as
$$
v_{x,S}^\Phi(\br) = - \sum_{i,j=1}^N \frac{\phi_i(\br)\phi_j(\br)}{\rho_\Phi(\br)}
\int_{\RR^3} \frac{\phi_i(\br')\phi_j(\br')}{|\br-\br'|} \, d\br',
$$
remark that
$$
\left| \frac{\phi_i(\br)\phi_j(\br)}{\rho_\Phi(\br)} \right| \le 1,
$$
and conclude using the following Lemma. 

\medskip

\begin{lemme} \label{Lem:Coulomb}
Let $\Phi = (\phi_i)_{1 \le i \le N}\in {\cal X}_N$ and
$$
V_{ij}(\br) = \int_{\RR^3} \frac{\phi_i(\br') \, \phi_j(\br')}{|\br-\br'|} \, d\br'.
$$
Then $V_{ij}$ vanishes at infinity. Besides, if the $\phi_i$ are radial
or if there exists $1 \le p < 3/2 < q \le 2$ such that $|\br| \, |\phi_i\phi_j|
    \in L^p(\RR^3) \cap L^q(\RR^3)$, then
$$
V_{ij}(\br) = \frac{\delta_{ij}}{|\br|} + o\left( \frac{1}{|\br|} \right).
$$
\end{lemme}

\medskip

\noindent
{\it Proof of Lemma~\ref{Lem:Coulomb}.} Let us denote by $\rho_{ij} =
\phi_i\phi_j$. By Sobolev embeddings, $\rho_{ij} \in L^1(\RR^3)\cap
L^3(\RR^3)$. For all $R > 0$ and all $\br \in \RR^3$ such that $|\br| \ge
2R$, one has
$$
|V_{ij}(\br)| \le \int_{|\br'| < R} \frac{|\rho_{ij}(\br')|}{|\br-\br'|} \, d\br' + 
\int_{|\br'| > R} \frac{|\rho_{ij}(\br')|}{|\br-\br'|} \, d\br 
\le \frac 1 R + \left\| |\rho_{ij}| \chi_{B_R^c} \star \frac{1}{|\cdot|}
\right\|_{L^\infty}. 
$$
It then follows from the Young inequality and the Lebesgue dominated
convergence theorem that
\begin{eqnarray*}
 \left\| |\rho_{ij}| \chi_{B_R^c} \star \frac{1}{|\cdot|}
 \right\|_{L^\infty} & \le &
 \left\| |\rho_{ij}| \chi_{B_R^c} \star \frac{\chi_{B_1}}{|\cdot|}
 \right\|_{L^\infty} + 
 \left\| |\rho_{ij}| \chi_{B_R^c} \star \frac{\chi_{B_1^c}}{|\cdot|}
 \right\|_{L^\infty} \\
& \le & \| |\rho_{ij}| \chi_{B_R^c} \|_{L^3} \, 
\left\|  \frac{\chi_{B_1}}{|\cdot|} \right\|_{L^{3/2}} +
\| |\rho_{ij}| \chi_{B_R^c} \|_{L^1} \, 
\left\|  \frac{\chi_{B_1^c}}{|\cdot|} \right\|_{L^{\infty}} \;
\mathop{\longrightarrow}_{R \to +\infty} 0.
\end{eqnarray*}
Therefore, $V_{ij}$ vanishes at infinity.

\medskip

\noindent
The case of radial orbitals can be dealt with using the Gauss theorem, which
provides the following expression for the potential $V_{ij}$:
$$
 V_{ij}(\br) = \int_\mR \frac{\rho_{ij}(\br')}{\max(|\br|,|\br'|)} \, d\br' \qquad
 \mbox{(radial orbitals)}.
$$
Indeed,
\begin{eqnarray}
\left| V_{ij}(\br) - \frac{\delta_{ij}}{|\br|} \right| & = &
\left| \int_\mR \frac{\rho_{ij}(\br')}{\max(|\br|,|\br'|)} \, d\br' 
- \frac{\delta_{ij}}{|\br|} \right| 
=  \left| - \frac{1}{|\br|} \int_{|\br'| \geq |\br|} \rho_{ij} 
+ \int_{|\br'| \geq |\br|} \frac{\rho_{ij}(\br')}{|\br'|} \, d\br' \right| \nonumber \\ 
& \le & \frac{2}{|\br|} \, \int_{|\br'| \geq |\br|} |\rho_{ij}(\br')| \, d\br'.
\label{eq:bound_Vij}
\end{eqnarray}
We conclude using Lebesgue dominated convergence theorem.

\medskip

\noindent
Let us now prove~(\ref{eq:decay_vxS}) in the general case (non-radial
orbitals), under the additional assumption that there exists $1 \le p <
3/2 < q \le 2$ such that $|\br| \, |\rho_{ij}| \in L^p(\RR^3) \cap
L^q(\RR^3)$. For all $\br \in \RR^3$, 
$$
\left| |\br|V_{ij}(\br) - \delta_{ij} \right| = 
\left| \int_{\RR^3} \frac{|\br|-|\br-\br'|}{|\br-\br'|}  \rho_{ij}(\br') \,
  d\br' \right|  \le   \int_{\RR^3}  \frac{|\br'| \, |\rho_{ij}(\br')|}{|\br-\br'|}
 \,  d\br'.
$$
It suffices to show that the right-hand side vanishes at infinity.
For all $R > 0$ et all $\br \in \RR^3$ such that $|\br| \ge R(R+1)$,
\begin{eqnarray*}
\int_{\RR^3}  \frac{|\br'| \, |\rho_{ij}(\br')|}{|\br-\br'|}
 \,  d\br' & = & \int_{|\br'|<R}  \frac{|\br'| \, |\rho_{ij}(\br')|}{|\br-\br'|}
 \,  d\br' + \int_{|\br'|>R}  \frac{|\br'| \, |\rho_{ij}(\br')|}{|\br-\br'|}
 \,  d\br' \\ & \le & \frac 1 R + \left\| f_{ij} \chi_{B_R^c} \star
   \frac{1}{|\cdot|} \right\|_{L^\infty},
\end{eqnarray*}
where $f_{ij}(\br) = |\br| \,  |\rho_{ij}(\br)|$. We then use the same
argument as above: 
\begin{eqnarray*}
\left\| f_{ij} \chi_{B_R^c} \star \frac{1}{|\cdot|} \right\|_{L^\infty}
& \le & 
 \left\| f_{ij} \chi_{B_R^c} \star \frac{\chi_{B_1}}{|\cdot|}
 \right\|_{L^\infty} + 
 \left\| f_{ij} \chi_{B_R^c} \star \frac{\chi_{B_1^c}}{|\cdot|}
 \right\|_{L^\infty} \\
& \le & \| f_{ij} \chi_{B_R^c} \|_{L^q} \, 
\left\|  \frac{\chi_{B_1}}{|\cdot|} \right\|_{L^{q'}} +
\| f_{ij} \chi_{B_R^c} \|_{L^p} \, 
\left\|  \frac{\chi_{B_1^c}}{|\cdot|} \right\|_{L^{p'}} \;
\mathop{\longrightarrow}_{R \to +\infty} 0,
\end{eqnarray*}
where $p'=(1-p^{-1})^{-1} \in (3,+\infty]$ and $q'=(1-q^{-1})^{-1} \in
[2,3)$.
The proof of Lemma~\ref{Lem:Coulomb} is complete. 
\cqfd

\medskip

\noindent
Let us now turn to the proof of the second assertion of
Proposition~\ref{prop:quali_Slater}. For all $v \in L^3(\RR^3) +
L^\infty(\RR^3)$, the operator $v \gamma_\Phi$ is
Hilbert-Schmidt. Indeed
$$
(v \gamma_\Phi)(\br,\br') = v(\br) \gamma_\Phi(\br,\br')  \in L^2(\RR^3 \times \RR^3)
$$
since $|\gamma_\Phi(\br,\br')| \le \rho_{\Phi}(\br) \, \rho_{\Phi}(\br')$ with 
$\rho_\Phi \in L^1(\RR^3) \cap L^3(\RR^3)$.
One can thus define on $L^3(\RR^3) + L^\infty(\RR^3)$ the functional
$$
J_\Phi^S(v) = \frac 1 2 \| v \gamma_\Phi- K_\Phi \|_\HS^2 =
\frac 1 2 \int_{\RR^3} \int_{\RR^3} \left
| v(\br) \gamma_\Phi(\br,\br')  +
  \frac{\gamma_\Phi(\br,\br')}{|\br-\br'|} \right|^2 \, d\br \, d\br'.  
$$
For all $v$ and $h$ in $L^3(\RR^3) + L^\infty(\RR^3)$, 
$$
J_\Phi^S(v+h) = J_\Phi^S(v) + 
\int_{\RR^3} \left( v(\br) \rho_\Phi(\br) + \int_{\RR^3}
  \frac{|\gamma(\br,\br')|^2}{|\br-\br'|}  \, d\br' \right) \, h(\br) \, d\br
+ \frac 1 2  \| h \gamma_\Phi  \|_\HS^2.
$$
Therefore, all the local minima of $J_\Phi^S$ are global, and they are
characterized by the equation
$$
v(\br) \rho_\Phi(\br) + \int_{\RR^3}
  \frac{|\gamma(\br,\br')|^2}{|\br-\br'|}  \, d\br' = 0.
$$
If $\rho_\Phi > 0$ almost everywhere, the Slater potential is the unique
solution to the above equation, and therefore the unique global minimizer
of $J_\Phi^S$.

The fact that the minimizers of $J_\Phi^S$ and $I_\Phi^S$ are
the same comes from the fact that $\gamma_\Phi^2 = \gamma_\Phi$
implies $\langle  K_\Phi \gamma_\Phi,  v \gamma_\Phi
\rangle_\HS = \langle K_\Phi,  v \gamma_\Phi \rangle_\HS$. 
\cqfd

\subsection{Proof of Theorem~\ref{thm:fixed_point_Slater}:
  Self-consistent Slater equation} 

\noindent
The strategy of proof is based on a fixed-point argument. Notice that
variational methods cannot be used since (\ref{eq:SCF_Slater}) seems to
have no variational interpretation.  

\medskip

\noindent
For all $\eta \ge  0$, we consider the problem 
\begin{eqnarray} \label{eq:sys_FP}
\left\{ \begin{array}{l}
\dps \left( -\frac12 \Delta - \frac{Z+\eta}{|\br|} + \rho_{\Phi^\eta}
\star \frac{1}{|\br|} + v^{\Phi^\eta,\eta}_\xS  \right) \phi_i^\eta = \epsilon_i^\eta
\phi_i^\eta, \\
\dps \int_{\RR^3} \phi_i^\eta \phi_j^\eta = \delta_{ij}, \\
\epsilon_1^\eta \le \cdots \le \epsilon_N^\eta \mbox{ are the lowest $N$
  eigenvalues of } \left( -\frac12 \Delta - \frac{Z+\eta}{|\br|} +
  \rho_{\Phi^\eta} 
\star \frac{1}{|\br|} + v^{\Phi^\eta,\eta}_\xS  \right) \mbox{ (on
$L^2_r(\RR^3)$)}
\end{array} \right.
\end{eqnarray}
where 
$$
v^{\Phi,\eta}_\xS(\br) = - \frac{1}{\rho_\Phi(\br)+\eta} \int_{\RR^3}
\frac{|\gamma_\Phi(\br,\br')|^2}{|\br-\br'|} \, d\br'. 
$$ 
The proof of existence of a solution to (\ref{eq:sys_FP}) for $\eta = 0$
follows the lines of the proof of Theorem~III.3
in~\cite{Lions87}. We first construct, for $\eta > 0$, a continuous
application $T^\eta$ 
whose fixed points are solutions to
(\ref{eq:sys_FP}) in ${\cal X}_N^r$. We then prove the existence of a
fixed point of $T^\eta$ using 
Schauder Theorem. The existence of a solution to (\ref{eq:sys_FP}) in
the case when $\eta = 0$ is finally obtained using some
limiting procedure. Note that we have introduced the parameter $\eta$
both in the nucleus-electron interaction and in the Slater potential. In
the former term, $\eta$ plays the same role as in~\cite{Lions87}
(i.e. it enables us to control the decay of the orbitals
at infinity). The role of $\eta$ in the latter term is to ensure the
continuity of the nonlinear application $T^\eta$ for $\eta > 0$.

\medskip

\noindent
{\it First step.} Construction of the application $T^\eta$.

\medskip

\noindent
Let $\eta > 0$ and
$$
K = \left\{ \Psi = (\psi_i)_{1 \le i \le N} \in (H^1_r(\RR^3))^N \;
\left | \; \left[ \int_{\RR^3} \psi_i \psi_j \right] \le I_N \right. \right\},
$$
$I_N$ denoting the identity matrix of rank $N$. The semidefinite
constraint $\left[ \int_{\RR^3} \psi_i \psi_j \right] \le I_N$ means
$$
\forall {\bold x} \in \RR^N, \quad \sum_{i,j=1}^N \left(\int_{\RR^3} \psi_i
  \psi_j \right) x_i x_j \le |{\bold x}|^2.
$$
It is easy to see that $K$ is a nonempty, closed, bounded, convex subset
of the Hilbert space $(H^1_r(\RR^3))^N$, containing ${\cal X}_N^r$. For
$\Psi \in K$, we denote by $\gamma_\Psi(\br,\br') = \sum_{i=1}^N
\psi_i(\br) \psi_i(\br')$, $\rho_\Psi(\br) = \gamma_\Psi(\br,\br)$ and
$$
\widetilde F_\Psi^\eta = -\frac12 \Delta - \frac{Z+\eta}{|\br|} + \rho_{\Psi}
\star \frac{1}{|\br|} + v^{\Psi,\eta}_\xS.
$$
As the potential $V_\Psi^\eta = - \frac{Z+\eta}{|\br|} + \rho_{\Psi}
\star \frac{1}{|\br|} + v^{\Psi,\eta}_\xS$ belongs to 
$$
L^2(\RR^3) + L^\infty_\epsilon(\RR^3) = \left\{ W \; | \; \forall
  \epsilon > 0, \; \exists (W_2,W_\infty) \in L^2(\RR^3) \times
  L^\infty(\RR^3), \; \|W_\infty\|_{L^\infty} \le \epsilon, \; W =
  W_2+W_\infty \right\},
$$
it is a compact perturbation of the kinetic energy operator. By Weyl
Theorem~\cite{ReedSimon}, $\sigma_{\rm ess}(\widetilde F_\Psi^\eta) = 
\sigma_{\rm ess}(-\frac 12 \Delta) = [0,\infty)$. Besides, using Gauss
theorem and the inequalities $- \frac{N}{|\cdot|} \le - \rho_{\Psi}
\star \frac{1}{|\br|} \le 
v^{\Psi,\eta}_\xS \le 0$, one has $- \frac{Z+\eta}{|\br|} \le V_\Psi^\eta \le -
\frac{\eta}{|\br|}$. Hence, 
\begin{equation} \label{eq:Hlike}
 {\cal G}^{Z+\eta} := -\frac12 \Delta -
\frac{Z+\eta}{|\br|} \le \widetilde F_\Psi^\eta \le {\cal G}^\eta :=
-\frac12 \Delta - \frac{\eta}{|\br|}. 
\end{equation}
As the hydrogen-like Hamiltonian ${\cal G}^\eta$,
considered as an operator on 
$L^2_r(\RR^3)$, has infinitely many negative eigenvalues, so does  
$\widetilde F_\Psi^\eta$ (this is a straightforward consequence of
Courant-Fischer min-max principle). Besides, the eigenvalues of 
the radial Schrödinger operator $\widetilde F_\Psi^\eta$ being
simple, the spectral problem
\begin{eqnarray*} 
\left\{ \begin{array}{l}
\dps  \widetilde F_\Psi^\eta \phi_i = \epsilon_i \phi_i, \\
\dps \int_{\RR^3} \phi_i \phi_j = \delta_{ij}, \\
\epsilon_1 \le \cdots \le \epsilon_N \mbox{ are the lowest $N$
  eigenvalues of } \widetilde F_\Psi^\eta  \mbox{ (on
$L^2_r(\RR^3)$)},
\end{array} \right.
\end{eqnarray*}
has a unique solution $\Phi = (\phi_i)$ in ${\cal X}_N^r \subset K$ up to the
signs of the orbitals $\phi_i$. We can therefore define a nonlinear
application $T^\eta$ from $K$ to $K$ which
associates with any $\Psi \in K$ the unique solution $\Phi =
(\phi_i) \in {\cal X}_N^r \subset K$ to (\ref{eq:sys_FP}), for which
$\phi_i \ge 0$ in a neighborhood of $\br=0$, for all $1 \le i \le N$ (by
the strong maximum principle, $\phi_i$ cannot vanish on an open set of
$\RR^3$).  

\medskip

\noindent
{\it Second step.} Existence of a solution to (\ref{eq:sys_FP}) for 
$\eta > 0$. 

\medskip

\noindent
Using standard perturbation arguments, it
is not difficult to prove that $T^\eta$ is continuous (for the
$H^1$ norm topology). Let us prove that $T^\eta$ is compact. Let
$(\Psi^n)$ be a bounded sequence in $K$, and let $\Phi^n =
T^\eta \Psi^n$. There is no restriction in assuming that $(\Psi^n)$
converges to some $\Psi^\eta \in (H^1(\RR^3))^N$, weakly in
$(H^1(\RR^3))^N$, strongly in $(L^2_{\rm loc}(\RR^3))^N$ and almost
everywhere. This implies in particular that the sequence $(\rho_{\Psi^n}
\star \frac{1}{|\br|} + v_\xS^{\Psi^n,\eta})$ is bounded in $L^\infty$ and
converges almost everywhere to $\rho_{\Psi^\eta}
\star \frac{1}{|\br|} + v_\xS^{\Psi^\eta,\eta}$ when $n$ goes to infinity.
Using again (\ref{eq:Hlike}) and denoting by
$\epsilon_i^n$ the $i$-th eigenvalue of $\widetilde F_{\Psi^n}^\eta$,
one obtains  
$$ 
\frac 1 2 \sum_{i=1}^N \left( \| \nabla \phi_i^n \|_{L^2} - 2(Z+\eta)
\right)^2 - 2 (Z+\eta)^2 \le 
\sum_{i=1}^N \frac 1 2 \int_{\RR^3} |\nabla \phi_i^n|^2 - 
\int_{\RR^3} \frac{Z+\eta}{|\br|} \rho_{\Phi^n} \le \sum_{i=1}^N
\epsilon_i^n < 0.
$$
Thus, for all $1 \le i \le N$, the sequence $(\phi_i^n)_{n \in
  \NN^\ast}$ is uniformly bounded in $H^1(\RR^3)$ (independently of
$(\Psi^n)$), and therefore converges, up to extraction, to some
$\phi_i^\eta \in H^1_r(\RR^3)$, weakly in $H^1(\RR^3)$, strongly in
$L^2_{\rm loc}(\RR^3)$ and almost everywhere.
Besides, using (\ref{eq:Hlike}) and Courant-Fischer formula, one obtains
$$
- \frac{(Z+\eta)^2}{2 i^2} \le \epsilon_i^n \le - \frac{\eta^2}{2 i^2}.
$$
Up to extraction, $(\epsilon_i^n)$ therefore converges to some 
$\epsilon_i^\eta \in [- \frac{(Z+\eta)^2}{2 i^2},- \frac{\eta^2}{2 i^2}]$.
Next, by Kato inequality~\cite{ReedSimon}, 
\begin{eqnarray}
  \label{eq:boundPhi}
  - \Delta |\phi_i^n| & \le & - \mbox{sgn}(\phi_i^n) \Delta
  \phi_i^n = 2 (\epsilon_i^n - V^\eta_{\Psi^n}) |\phi_i^n| \nonumber \\
  & \le & 2 \left( \frac{Z+\eta}{|\br|} - \frac{\eta^2}{i^2} \right)
  |\phi_i^n|. 
\end{eqnarray}
As, moreover, $(\Psi^n)$ and $(\Phi^n)$ are bounded for the $H^1$ norm
topology, $(V^\eta_{\Psi^n} \phi_i^n)$ is bounded in $L^2(\RR^3)$, so that 
$(\phi_i^n)$ is bounded in $H^2(\RR^3)$, hence in
$L^\infty(\RR^3)$. Consequently, it follows from (\ref{eq:boundPhi}) and
the maximum principle that there exists $\delta > 0$ small enough and $M \ge 0$
independent of~$i$ and~$n$, such that
$$
|\phi_i^n(\br)| \le M \, {\rm e}^{-\left( \frac{\sqrt 2 \, \eta}{N} -
  \delta \right) |\br|}.
$$ 
This implies that $(\phi_i^n)_{n \in \NN^\ast}$ converges  (up to 
extraction) to $\phi_i^\eta$  strongly in 
$L^2(\RR^3)$. In particular, $\Phi^\eta = (\phi_i^\eta) \in {\cal 
  X}_N^r$. It is then possible to check, using the convergence of
$(\Psi^n)$ to $\Psi^\eta$ and the convergence - up to extraction - of
$(\Phi^n)$ to $\Phi^\eta$ and of $(\epsilon_i^n)$ to $\epsilon_i^\eta$,
that 
$$
- \frac 1 2 \Delta \phi_i^\eta + V_{\Psi^\eta}^\eta \phi_i^\eta =
\epsilon_i^\eta \phi_i^\eta
$$
and next, using the positivity of $\rho_{\Psi^n}
\star \frac{1}{|\br|} + v_\xS^{\Psi^n,\eta}$ and Fatou lemma, that
\begin{eqnarray*}
\liminf_{n \to +\infty} - \int_{\RR^3} |\nabla \phi_i^n|^2 & = &
\liminf_{n \to +\infty} 2 \int_{\RR^3}
(V_{\Psi^n}^\eta-\epsilon_i^n) |\phi_i^n|^2 \\
& \ge & 2 \int_{\RR^3}
(V_{\Psi^\eta}^\eta-\epsilon_i^\eta) |\phi_i^\eta|^2 = - \int_{\RR^3}
|\nabla \phi_i^\eta|^2. 
\end{eqnarray*}
As on the other hand, 
$$
\int_{\RR^3} |\nabla \phi_i^\eta|^2 \le \liminf_{n \to +\infty} 
\int_{\RR^3} |\nabla \phi_i^n|^2, 
$$
$(\Psi^n)$ converges to $\Psi^\eta$ strongly in
$(H^1(\RR^3))^N$, which proves that $T^\eta$ is compact. It then
follows from Schauder fixed point theorem~\cite{Zeidler} that
$T^\eta$ has a fixed point $\Phi^\eta \in {\cal X}_N^r$, which is
solution to (\ref{eq:sys_FP}).

\medskip

\noindent
{\it Third step.} Existence of a solution to (\ref{eq:sys_FP}) for 
$\eta = 0$. 

\medskip

\noindent
Let $(\eta_n)$ be a sequence of positive real numbers converging to
zero. As the sequence of corresponding fixed points $(\Phi^{\eta_n})$ is
uniformly bounded in $(H^1(\RR^3))^N$ and as $- \frac{(Z+\eta_n)^2}{2i^2}
\le \epsilon_i^{\eta_n} \le 0$, there is no restriction in 
assuming that $(\Phi^{\eta_n})$ converges to some 
$\Phi^\star \in (H^1(\RR^3))^N$, weakly in $(H^1(\RR^3))^N$, strongly in
$(L^2_{\rm loc}(\RR^3))^N$ and almost everywhere, and that
$(\epsilon_i^{\eta_n})$ converges to $\epsilon_i^\ast \le 0$. Besides,
the sequence $(\Phi^{\eta_n})$ is bounded in $(H^2(\RR^3))^N$, hence in 
$(L^\infty(\RR^3))^N$.

\medskip

\noindent
Passing to the limit in the equation $\widetilde
  F_{\Phi^{\eta_n}}^{\eta_n} \phi_i^{\eta_n} =
\epsilon_i^{\eta_n}\phi_i^{\eta_n}$ yields
$$
- \frac 1 2 \Delta \phi_i^\star - \frac{Z}{|\br|} \phi_i^\star + \left(
  \rho_{\Phi^\star} \star \frac{1}{|\br|}\right)  \phi_i^\star +
v_\xS^{\Phi^\star} \phi_i^\star =
\epsilon_i^\star \phi_i^\star.
$$

\noindent
Assume that $\int_{\RR^3} \rho_{\Phi^\star} < N$. As
$$
\widetilde F_{\Phi^{\eta_n}}^{\eta_n} \le 
-\frac 1 2 \Delta - \frac{Z}{|\br|} + \rho_{\Phi^{\eta_n}} \star \frac{1}{|\br|},
$$
one has, using Courant-Fischer formula, and denoting by $\lambda_i(A)$
the $i$-th eigenvalue of $A$, 
\begin{eqnarray*}
\epsilon_i^\star & = & \lim_{n \to +\infty}  \epsilon_i^{\eta_n} \\
& = & \lim_{n \to +\infty}  \lambda_i \left( \widetilde 
    F_{\Phi^{\eta_n}}^{\eta_n} \right) \\
& \le  & \lim_{n \to +\infty} \lambda_i \left( 
-\frac 1 2 \Delta - \frac{Z}{|\br|} + \rho_{\Phi^{\eta_n}} \star
\frac{1}{|\br|} \right) \\
& = &  \lambda_i \left( 
-\frac 1 2 \Delta - \frac{Z}{|\br|} + \rho_{\Phi^{\star}} \star
\frac{1}{|\br|} \right) \\
& \le &  \lambda_i \left( 
-\frac 1 2 \Delta - \frac{N- \int_{\RR^3} \rho_{\Phi^\star}}{|\br|}
\right) \\
& = & - \frac{(N- \int_{\RR^3} \rho_{\Phi^\star})^2}{2i^2} < 0.
\end{eqnarray*}
It follows that for $n$ large enough, the sequence
$(\epsilon_i^{\eta_n})$ is isolated from zero. As $(\Phi^{\eta_n})$ is
bounded in $(L^\infty(\RR^3))^N$, we conclude, reasoning as above, that
there exists $M \in \RR_+$ and $\alpha > 0$ such that for $n$ large enough
$$
|\phi_i^{\eta_n}(\br)| \le M \, {\rm e}^{-\alpha |\br|}.
$$
This implies that  $(\Phi^{\eta_n})$ converges to
$\Phi^\star \in (H^1(\RR^3))^N$ strongly in
$(L^2(\RR^3))^N$, and consequently that $\int_{\RR^3} \rho_{\Phi^\star} = N$. We
reach a contradiction. This means that $\int_{\RR^3} \rho_{\Phi^\star} =
N$ and therefore that $\Phi^\star \in {\cal X}_N^r$.

\medskip

\noindent
This proves that $(\phi_i^\ast)$ are orthonormal eigenvectors of
$\widetilde F_{\Phi^\star}^0$. The fact that $\epsilon_1^\star < \cdots
< \epsilon_N^\star$ are the lowest eigenvalues of $\widetilde
F_{\Phi^\star}^0$ follows from Courant-Fischer formula.

\medskip

\noindent
In view of Proposition~\ref{prop:quali_Slater}, the Slater potential
$v_{x,S}^{\Phi^\star}$ is equivalent to $-\frac{1}{|\br|}$ at
infinity. This proves that $\epsilon_1^\star < \cdots < \epsilon_N^\star
< 0$, from which it follows that the orbitals $\phi_i^\star$ enjoy
exponential decay: For all $\eta > 0$, there exists $M \in \RR^3$ such
that
$$
|\phi_i^\star(\br)| \le M \, {\rm e}^{- (\sqrt{-2\epsilon_N^\star}-\eta/3)|\br|}.
$$
Using (\ref{eq:bound_Vij}), one obtains
$$
v^{\Phi^\star}_\xS(\br) = -\frac{1}{|\br|} + {\rm
  o}\left({\rm e}^{- (2 \, \sqrt{-2\epsilon_N^\star}-\eta)|\br|}\right).
$$

\medskip

\noindent
Lastly, the same arguments as above can be used to prove that the
minimum of the Hartree-Fock energy over the set of
solutions to~(\ref{eq:SCF_Slater})  
is attained.
\cqfd


\subsection{Proof of Theorem~\ref{thm:integral_OEP}: OEP Integral equation} 
\label{sec:proof_integral_OEP}

Straightforward computations show that
\begin{equation}
\label{eq:difference_HF_OEP}
{\cal E}^{\rm HF}(\gamma_{W+w}) =  {\cal E}^{\rm HF}(\gamma_{W}) + \Tr({\cal
  F}_{\gamma_W} (\gamma_{W+w} - \gamma_W)) 
+ \alpha(\gamma_{W+w}-\gamma_{W},\gamma_{W+w}-\gamma_{W}),
\end{equation}
where 
$$
\alpha(\gamma_1,\gamma_2) = \frac12 \int_\mR \int_\mR \frac{\gamma_1(\br,\br)
  \, \gamma_2(\br',\br') - \gamma_1(\br,\br') \, \gamma_2(\br,\br')}{|\br-\br'|} \,
d\br \, d\br'. 
$$
The second term of the right-side of (\ref{eq:difference_HF_OEP}) is
well-defined since both $\gamma_{W+w}$ and $\gamma_W$ are finite rank
operators with range in $H^2(\RR^3) = D({\cal F}_{\gamma_W}) = D({\cal F}_{\gamma_{W+w}})$. 
Let
$$
\delta = \left( \mathop{\mbox{max}}_{z \in {\cal C}} \|R^0(z)\|
\right)^{-1}. 
$$
Denoting by $R^w(z) = (z-H_{W+w})^{-1}$, one has for all $w \in
{\cal B}_{\eta/2} = \left\{ w \in L^1(\RR^3) \cap L^\infty(\RR^3), \;  
\| w \|_{L^1 \cap L^\infty} < \eta/2 \right\}$ such that
$\|w\|_{L^\infty} < \delta$, 
$$
R^w(z) = (z-H_{W+w})^{-1} = (z-H_W - w)^{-1} = ((z-H_W)(1-R^0(z)w))^{-1}
= (1-R^0(z)w)^{-1} R^0(z)
$$
and 
$$
(1-R^0(z)w)^{-1}-1 = (1-R^0(z)w)^{-1} R^0(z) w = R^0(z) w (1-R^0(z)w)^{-1} .
$$
Using the complex-plane integral representation
$$
\gamma_{W+w} = \frac{1}{2\pi \ri} \oint_{\cal C} R^w(z)  \, dz,
$$
one is led to 
\begin{eqnarray*}
\gamma_{W+w} - \gamma_W  & = &  \frac{1}{2\pi \ri} \oint_{\cal C} (R^w(z)
-R^0(z))  \, dz   =   \frac{1}{2\pi \ri} \oint_{\cal C} (1-R^0(z)
w)^{-1} R^0(z) w 
R^0(z)  \, dz   \\ &  = &   \frac{1}{2\pi \ri} \oint_{\cal C} R^0(z)
 w R^0(z)  \, dz +  \frac{1}{2\pi \ri} \oint_{\cal C} 
R^0(z) w (1-R^0(z) w)^{-1}  R^0(z) w R^0(z)  \, dz . 
\end{eqnarray*}
Hence,
\begin{eqnarray*}
{\cal F}_{\gamma_W}(\gamma_{W+w} - \gamma_W)  & = & \frac{1}{2\pi \ri}
\oint_{\cal C} {\cal 
  F}_{\gamma_W} R^0(z)
 w R^0(z)  \, dz +  \frac{1}{2\pi \ri} \oint_{\cal C} 
{\cal F}_{\gamma_W} R^0(z) w (1-R^0(z) w)^{-1}  R^0(z) w R^0(z)  \, dz
\\
& = & \frac{1}{2\pi \ri} \oint_{\cal C} H_W R^0(z) w R^0(z)  \, dz + 
 \frac{1}{2\pi \ri} \oint_{\cal C} (K_{\gamma_W}-v_x^W) R^0(z) w R^0(z)
 \, dz
\\ & & +  \frac{1}{2\pi \ri} \oint_{\cal C} 
{\cal F}_{\gamma_W} R^0(z) w (1-R^0(z) w)^{-1}  R^0(z) w R^0(z)  \, dz \\
& = & \frac{1}{2\pi \ri} \oint_{\cal C} (-1+ z R^0(z)) w R^0(z)  \, dz + 
 \frac{1}{2\pi \ri} \oint_{\cal C} (K_{\gamma_W}-v_x^W) R^0(z) w R^0(z)
 \, dz
\\ & & +  \frac{1}{2\pi \ri} \oint_{\cal C} 
{\cal F}_{\gamma_W} R^0(z) w (1-R^0(z) w)^{-1}  R^0(z) w R^0(z)  \, dz
\\
& = & - w \gamma_W + \frac{1}{2\pi \ri} \oint_{\cal C} z R^0(z) w R^0(z)  \, dz + 
 \frac{1}{2\pi \ri} \oint_{\cal C} (K_{\gamma_W}-v_x^W) R^0(z) w R^0(z)
 \, dz
\\ & & +  \frac{1}{2\pi \ri} \oint_{\cal C} 
{\cal F}_{\gamma_W} R^0(z) w (1-R^0(z) w)^{-1}  R^0(z) w R^0(z)  \, dz.
\end{eqnarray*}

To proceed further, we make use of the following technical Lemmas, whose
proofs are postponed until the end of the present section.

\medskip

\begin{lemme}
\label{lem:resolvent}
For all $z \in \rho(H_W)$, $(1-\Delta) R^0(z)$ and
$R^0(z)(1-\Delta)$ are bounded operators on $L^2(\RR^3)$ and
$(1-\Delta)R^0(z)$ is the adjoint of $R^0(z)(1-\Delta)$. Besides the functions 
$$
z \mapsto R^0(z) (1-\Delta) \qquad \mbox{and} \qquad 
z \mapsto (1-\Delta) R^0(z)
$$ 
are analytic from $\rho(H_W)$ into ${\cal L}(L^2(\RR^3))$.
\end{lemme}

\medskip

\begin{lemme}
\label{lem:traceclass} $\,$
\begin{enumerate}[(1)]
\item 
For all  $v \in L^1(\RR^3)$, the operator 
$(1-\Delta)^{-1}v(1-\Delta)^{-1}$ is trace-class and
$$
\|(1-\Delta)^{-1}v(1-\Delta)^{-1}\|_{\tc} \le \frac{1}{8\pi} \,
\|v\|_{L^1}. 
$$
\item For all $v \in L^2(\RR^3)$, the operator $v(1-\Delta)^{-1}$ and
  its adjoint $(1-\Delta)^{-1}v$ are Hilbert-Schmidt and 
$$
\| v(1-\Delta)^{-1} \|_\HS = \| (1-\Delta)^{-1}v \|_\HS =
\frac{\|v\|_{L^2}}{(8\pi)^{1/2}}.
$$ 
\end{enumerate}
\end{lemme}

\medskip

\noindent
Using the above two Lemmas, it follows
\begin{eqnarray*}
\Tr \left( \frac{1}{2\pi \ri} \oint_{\cal C} z R^0(z) w R^0(z)  \, dz
\right) & = & \frac{1}{2\pi \ri} \oint_{\cal C} z \Tr \left(R^0(z) w
R^0(z)\right) \, dz \\
& = &  \frac{1}{2\pi \ri} \oint_{\cal C} z \Tr \left( R^0(z) (1-\Delta)
(1-\Delta)^{-1} w (1-\Delta)^{-1} (1-\Delta) R^0(z)\right) \, dz \\ & =
& 
\frac{1}{2\pi \ri} \oint_{\cal C} z \Tr \left( (1-\Delta) R^0(z)^2 (1-\Delta)
(1-\Delta)^{-1} w (1-\Delta)^{-1} \right) \, dz \\ & = & 
\Tr \left( (1-\Delta) \left( \frac{1}{2\pi \ri} \oint_{\cal C} z
    R^0(z)^2\, dz \right) (1-\Delta)
(1-\Delta)^{-1} w (1-\Delta)^{-1} \right).
\end{eqnarray*}
Denoting by
$$
H_W = \int_{-\infty}^{+\infty} \lambda \, dP_\lambda
$$
the spectral decomposition of $H_W$, it holds
\begin{eqnarray*}
 \frac{1}{2\pi \ri} \oint_{\cal C} z R^0(z)^2\, dz & = & \frac{1}{2\pi \ri}
 \oint_{\cal C}  \left( \int_{-\infty}^{+\infty} \frac{z}{(z-\lambda)^2}
   \, dP_\lambda \right) \, dz \\
& = & \int_{-\infty}^{+\infty} \left( \frac{1}{2\pi \ri} \oint_{\cal C} 
\frac{z}{(z-\lambda)^2} \, dz \right)  dP_\lambda \\
& = & \int_{-\infty}^{\epsilon_F} dP_\lambda = \gamma_{W}.
\end{eqnarray*}
Hence,
$$
\Tr \left( \frac{1}{2\pi \ri} \oint_{\cal C} z R^0(z) w R^0(z)  \, dz
\right) = \Tr \left( (1-\Delta) \gamma_W (1-\Delta)
(1-\Delta)^{-1} w (1-\Delta)^{-1} \right) = \Tr(\gamma_W w).
$$
We thus obtain
\begin{eqnarray}
\label{eq:difference_HF_OEP_2}
{\cal E}^{\rm HF}(\gamma_{W+w}) & = & {\cal E}^{\rm HF}(\gamma_{W}) + \Tr \left(
  (K_{\gamma_W} - v_x^W)  \frac{1}{2\pi \ri} \oint_{\cal C} R^0(z) w 
R^0(z) \, dz \right)  \nonumber \\ & & + \Tr \left( 
\frac{1}{2\pi \ri} \oint_{\cal C} 
{\cal F}_{\gamma_W} R^0(z) w (1-R^0(z) w)^{-1}  R^0(z) w R^0(z)  \, dz
\right)
\\ & & + \alpha(\gamma_{W+w}-\gamma_{W},\gamma_{W+w}-\gamma_{W}). \nonumber
\end{eqnarray}
Let us denote by 
$$
\beta = \mathop{\mbox{max}}_{z \in {\cal C}} \|(1-\Delta)R^0(z)\|.
$$
As $W \in L^2(\RR^3)+L^\infty(\RR^3)$, one also has $v_x^W \in
L^2(\RR^3)+L^\infty(\RR^3)$. Let $v_2 \in L^2(\RR^3)$ and $v_\infty \in
L^\infty(\RR^3)$ such that $v_x^W = v_2 + v_\infty$. Then,
\begin{eqnarray*}
\left|\Tr \left(
(K_{\gamma_W} - v_x^W)   \frac{1}{2\pi \ri} \oint_{\cal C} R^0(z) w 
R^0(z) \, dz \right) \right| & \le &
\left|\Tr \left( (K_{\gamma_W} - v_\infty)
  \frac{1}{2\pi \ri} \oint_{\cal C}  R^0(z) w 
R^0(z) \, dz \right) \right| \\ & & + 
\left|\Tr \left( v_2
  \frac{1}{2\pi \ri} \oint_{\cal C} R^0(z) w 
R^0(z) \, dz \right) \right|,
\end{eqnarray*}
with
\begin{eqnarray}
\left| \Tr \left( (K_{\gamma_W} -
    v_\infty) \frac{1}{2\pi \ri} \oint_{\cal C}  R^0(z) w  R^0(z) \, dz
  \right) \right| 
& \le & \frac{|{\cal C}| \, \beta^2}{2\pi} 
(\|K_{\gamma_W}\|+ \|v_\infty\|_{L^\infty}) 
 \|(1-\Delta)^{-1} w (1-\Delta)^{-1}\|_{\tc} \nonumber \\
& \le & C \, \|w\|_{L^1} \le C \|w\|_{L^1 \cap L^\infty}, 
\label{eq:cont_diff1}
\end{eqnarray}
and
\begin{eqnarray}
\left|\Tr \left(  v_2 \frac{1}{2\pi \ri} \oint_{\cal C} R^0(z) w 
R^0(z) \, dz \right) \right| & = & 
\left|\Tr \left( \frac{1}{2\pi \ri} \oint_{\cal C}  v_2 (1-\Delta)^{-1}
    (1-\Delta)  R^0(z) 
    w (1-\Delta)^{-1} (1-\Delta) R^0(z) \, dz \right) \right| \nonumber \\
& \le & \frac{|{\cal C}| \, \beta^2}{2\pi} \, 
\| v_2 (1-\Delta)^{-1} \|_\HS \|w (1-\Delta)^{-1} \|_\HS \nonumber \\
& \le & C \, \|w\|_{L^2} \le C \|w\|_{L^1 \cap L^\infty}. 
\label{eq:cont_diff2}
\end{eqnarray}
The linear form
$$
w \mapsto \Tr \left( (K_{\gamma_W} - v_x^W)
  \frac{1}{2\pi \ri} \oint_{\cal C}  R^0(z) w 
R^0(z) \, dz \right) 
$$
therefore is continuous on $L^1(\RR^3) \cap L^\infty(\RR^3)$. It remains
to prove that the last two terms of the right-hand side of
(\ref{eq:difference_HF_OEP_2}) are ${\rm O}(\|w\|_{L^1\cap L^\infty}^2)$. The
first one is easy to deal with. Indeed,
\begin{eqnarray*}
& & \left|\Tr \left( 
\frac{1}{2\pi \ri} \oint_{\cal C} 
{\cal F}_{\gamma_W} R^0(z) w (1-R^0(z) w)^{-1}  R^0(z) w R^0(z)  \, dz
\right) \right|  \\
& \le & \frac{|{\cal
  C}|\beta^3  \|{\cal F}_{\gamma_W}(1-\Delta)^{-1}\|}
{2\pi\left(1-\frac{\|w\|_{L^\infty}}\delta\right)} \|w\|_{L^\infty} \, 
 \|(1-\Delta)^{-1} w (1-\Delta)^{-1}\|_{\tc} 
 \le   
\frac{|{\cal C}| \beta^3 \|{\cal F}_{\gamma_W}(1-\Delta)^{-1}\|}
{16\pi^2\left(1-\frac{\|w\|_{L^\infty}}\delta\right)} \|w\|_{L^1 \cap L^\infty}^2 .
\end{eqnarray*}
The second term can be split as
$$
\alpha(\gamma_{W+w}-\gamma_{W},\gamma_{W+w}-\gamma_{W})
= \frac 1 2 
D(\rho_{\gamma_{W+w}}-\rho_{\gamma_{W}},\rho_{\gamma_{W+w}}-\rho_{\gamma_{W}})
- \frac 1 2 
\int_{\RR^3}\int_{\RR^3}
\frac{|(\gamma_{W+w}-\gamma_W)(\br,\br')|^2}{|\br-\br'|} \, d\br \, d\br',
$$
where $D(\cdot,\cdot)$ denotes, as usual, the Coulomb energy
$$
D(f,g) = \int_{\RR^3}\int_{\RR^3} \frac{f(\br) \, g(\br')}{|\br-\br'|} \, d\br \, d\br',
$$
for which~\cite{ReedSimon}
$$
\exists C \in \RR_+ \mbox{ s.t. } \forall f \in L^{6/5}(\RR^3), \quad 
0 \le D(f,f) \le C \|f\|_{L^{6/5}}^2.
$$
As both $\rho_{\gamma_{W+w}}$ and $\rho_{\gamma_{W}}$ belong to $L^1(\RR^3)
\cap L^2(\RR^3)$, 
\begin{eqnarray*}
 D(\rho_{\gamma_{W+w}}-\rho_{\gamma_{W}},\rho_{\gamma_{W+w}}-\rho_{\gamma_{W}})
 & \le & C \,
\|\rho_{\gamma_{W+w}}-\rho_{\gamma_{W}}\|_{L^{6/5}}^2 \\
& \le & C \|\rho_{\gamma_{W+w}}-\rho_{\gamma_{W}}\|_{L^{1}}^{4/3} \,
\|\rho_{\gamma_{W+w}}-\rho_{\gamma_{W}}\|_{L^{2}}^{2/3}.
\end{eqnarray*}
We now make use of the following characterization of the $L^p$
norm~\cite{Lax02}, which is valid for all $1 \le p \le +\infty$: 
$$
\| f \|_{L^p} = \sup_{g \in L^{p'}(\mR), \;
  \|g\|_{L^{p'}}=1} \int_{\RR^3} f g,
$$
where $\frac{1}{p}+\frac{1}{p'} = 1$. In our case, one obtains
\begin{eqnarray*}
\|\rho_{\gamma_{W+w}}-\rho_{\gamma_{W}}\|_{L^{1}} & = & 
\sup_{g \in L^{\infty}, \; \|g\|_{L^{\infty}}=1} \int_{\RR^3}
(\rho_{\gamma_{W+w}}-\rho_{\gamma_{W}})  g \\ & = & 
\sup_{g \in L^{\infty}, \; \|g\|_{L^{\infty}}=1} \Tr(
(\gamma_{W+w}-\gamma_W)g ) \\
& \le & \sup_{g \in L^{\infty}, \; \|g\|_{L^{\infty}}=1}
\|(\gamma_{W+w}-\gamma_W)g \|_{\tc} \\
& = & \sup_{g \in L^{\infty}, \; \|g\|_{L^{\infty}}=1}
\left\|  \frac{1}{2\pi \ri} \oint_{\cal C} (1-R^0(z)
w)^{-1} R^0(z) w R^0(z) g  \, dz \right\|_{\tc} \\
& \le & 
\frac{|{\cal C}| \, \beta^2}{16\pi^2\left(1-\frac{\|w\|_{L^\infty}}\delta\right)}
\, \|w\|_{L^1},
\end{eqnarray*}
and
\begin{eqnarray*}
\|\rho_{\gamma_{W+w}}-\rho_{\gamma_{W}}\|_{L^{2}} & = & 
\sup_{g \in L^{2}, \; \|g\|_{L^{2}}=1} \int_{\RR^3}
(\rho_{\gamma_{W+w}}-\rho_{\gamma_{W}}) g
\\ & = & 
\sup_{g \in L^{2}, \; \|g\|_{L^{2}}=1} \Tr(
(\gamma_{W+w}-\gamma_W)g ) \\
& \le & \sup_{g \in L^{2}, \; \|g\|_{L^{2}}=1}
\|(\gamma_{W+w}-\gamma_W)g \|_{\tc} \\
& = & \sup_{g \in L^{2}, \; \|g\|_{L^{2}}=1}
\left\|  \frac{1}{2\pi \ri} \oint_{\cal C} (1-R^0(z)
w)^{-1} R^0(z) w R^0(z) g  \, dz \right\|_{\tc} \\
& = & \sup_{g \in L^{2}, \; \|g\|_{L^{2}}=1}
\frac{|{\cal C}| \, \beta^2}{2\pi(1-\frac{\|w\|_{L^\infty}}\delta)}
\|(1-\Delta)^{-1}w\|_\HS \, \|(1-\Delta)^{-1}g\|_\HS \\
& \le & \frac{|{\cal C}| \, \beta^2}{16\pi^2\left(1-\frac{\|w\|_{L^\infty}}\delta\right)}
\, \|w\|_{L^2}.
\end{eqnarray*}
Hence,
$$
0 \le D(\rho_{\gamma_{W+w}}-\rho_{\gamma_{W}},\rho_{\gamma_{W+w}}-\rho_{\gamma_{W}})
\le \frac{C}{1-\frac{\|w\|_{L^\infty}}\delta} \|w\|_{L^1\cap L^\infty}^2.
$$
Lastly, one obtains, using again Cauchy-Schwarz and Hardy inequalities,
$$
\int_{\RR^3}\int_{\RR^3}
\frac{|(\gamma_{W+w}-\gamma_W)(\br,\br')|^2}{|\br-\br'|} \, d\br \, d\br'
\le 2 \, \|\gamma_{W+w}-\gamma_W\|_\HS \, 
\| \nabla \gamma_{W+w}-\nabla \gamma_W \|_\HS.
$$
As
$$
\|\gamma_{W+w}- \gamma_W\|_\HS \le 
\|\gamma_{W+w}-\gamma_W\|_{\tc} \le  \frac{|{\cal C}| \,
  \beta^2}{16\pi^2\left(1-\frac{\|w\|_{L^\infty}}\delta\right)} 
\, \|w\|_{L^1}, 
$$
and
\begin{eqnarray*}
\|\nabla \gamma_{W+w}- \nabla \gamma_W\|_\HS & \le &
\bigg\| \nabla \frac{1}{2\pi \ri} \oint_{\cal C} R^0(z) w R^0(z) \, dz
\bigg\|_\HS \\ & & 
+ \bigg\| \nabla \frac{1}{2\pi \ri} \oint_{\cal C} R^0(z) w (1-R^0(z)
w)^{-1} R^0(z) w R^0(z) \, dz
\bigg\|_\HS \\ 
& & \hspace{-2cm} = \bigg\|  \frac{1}{2\pi \ri} \oint_{\cal C}
\nabla (1-\Delta)^{-1} (1-\Delta) R^0(z) w (1-\Delta)^{-1} (1-\Delta)
R^0(z) \, dz 
\bigg\|_\HS \\ & & \hspace{-1.9cm}  
+ \bigg\|  \frac{1}{2\pi \ri} \oint_{\cal C} \nabla (1-\Delta)^{-1}
(1-\Delta)  R^0(z) w (1-R^0(z)
w)^{-1} R^0(z) w (1-\Delta)^{-1} (1-\Delta) R^0(z) \, dz
\bigg\|_\HS \\ & \le & 
C \left( \|w\|_{L^2} + \frac{\|w\|_{L^2} \,
    \|w\|_{L^\infty}}{1-\frac{\|w\|_{L^\infty}}\delta} \right),
\end{eqnarray*}
we conclude that
$$
\int_{\RR^3} \int_{\RR^3}
\frac{|(\gamma_{W+w}-\gamma_W)(\br,\br')|^2}{|\br-\br'|} \, d\br \, d\br' 
= {\rm O}(\|w\|_{L^1\cap L^\infty}^2).
$$

\medskip

\noindent
We have therefore established that the Fr\'{e}chet derivative of the function
$w \mapsto {\cal E}^{\rm HF}(\gamma_{W+w})$ is the linear form
$$
w \mapsto  \Tr \left( (K_{\gamma_W} - v_x^W) 
  \frac{1}{2\pi \ri} \oint_{\cal C} R^0(z) w 
R^0(z) \, dz \right) .
$$
It follows from (\ref{eq:cont_diff1})-(\ref{eq:cont_diff2}) that this
linear form is in fact continuous on $L^1(\RR^3) \cap
L^2(\RR^3)$. Therefore, there exists $\varrho^W \in (L^1(\RR^3) \cap
L^2(\RR^3))' = L^2(\RR^3) + L^\infty(\RR^3)$, such that for all $w \in
L^1(\RR^3) \cap L^\infty(\RR^3)$,
$$
\Tr \left( (K_{\gamma_W} - v_x^W) 
  \frac{1}{2\pi \ri} \oint_{\cal C} R^0(z) w 
R^0(z) \, dz \right) = \int_{\RR^3} \varrho^W w.
$$
Using $[R^0(z),\gamma_W] = 0$, the analyticity of the
function $z \mapsto (1-\gamma_W)R^0(z)(1-\gamma_W)$ in the interior domain
defined by ${\cal C}$, and Cauchy's formula~\cite{Rudin}, it is easy to
show that 
\begin{eqnarray*}
\frac{1}{2\pi \ri} \oint_{\cal C} R^0(z) w 
R^0(z) \, dz & = & \frac{1}{2\pi \ri} \oint_{\cal C} 
 \gamma_W  R^0(z) \gamma_W w (1-\gamma_W)  R^0(z) (1-\gamma_W)  \, dz
\\ & & + \frac{1}{2\pi \ri} \oint_{\cal C} 
(1-\gamma_W)  R^0(z) (1-\gamma_W)   
w \gamma_W  R^0(z) \gamma_W \, dz.
\end{eqnarray*}
The left-hand side of the above equation therefore defines a finite-rank
operator. Let $(\phi_i^W)_{1 \le i \le N}$
be a set of $N$ orthonormal eigenvectors of $H_W$ associated with the
lowest $N$ eigenvalues $\epsilon_1^W \le \cdots \le \epsilon_N^W$ of
$H_W$. It holds
\begin{eqnarray*}
\frac{1}{2\pi \ri} \oint_{\cal C}  R^0(z) w 
R^0(z) \, dz & = & 
 \sum_{i=1}^N (|\phi_i^W\rangle \langle \phi_i^W|) w 
\frac{1}{2\pi \ri} \oint_{\cal C} \frac{1}{z-\epsilon_i^W}
 (1-\gamma_W)  R^0(z) (1-\gamma_W)  \, dz 
\\ & + & \sum_{i=1}^N \left( 
\frac{1}{2\pi \ri} \oint_{\cal C} \frac{1}{z-\epsilon_i^W}
 (1-\gamma_W)  R^0(z) (1-\gamma_W)  \, dz \right)   
w (|\phi_i^W\rangle \langle
 \phi_i^W|) .
\end{eqnarray*}
Using again  the analyticity of the
function $z \mapsto (1-\gamma_W)R^0(z)(1-\gamma_W)$ in the interior domain
defined by ${\cal C}$, and Cauchy's formula, we then obtain
$$
\frac{1}{2\pi \ri} \oint_{\cal C} \frac{1}{z-\epsilon_i^W}
 (1-\gamma_W)  R^0(z) (1-\gamma_W)  \, dz = 
 (1-\gamma_W)
  [\epsilon_i^W-(1-\gamma_W)H_W(1-\gamma_W)]^{-1}
 (1-\gamma_W) .
$$
Multiplying the above equality by $(K_{\gamma_W} - v_x^W)$ on the
left-hand side and taking the trace, we are led to 
$$
\Tr \left( (K_{\gamma_W} - v_x^W) 
  \frac{1}{2\pi \ri} \oint_{\cal C} R^0(z) w 
R^0(z) \, dz \right) = \int_{\RR^3} \varrho^W w
$$
with
$$
\varrho^W(\br) = 2 \, \sum_{i=1}^N \phi_i^W(\br) \; \left[ (1-\gamma_W)
  [\epsilon_i^W-(1-\gamma_W)H_W(1-\gamma_W)]^{-1}
 (1-\gamma_W)  (K_{\gamma_W} - v_x^W) \phi_i^W
\right] \!\! (\br).
$$
As the $\phi_i^W$s are in $H^2(\RR^3)$ and as the range of the operator
$[\epsilon_i^W-(1-\gamma_W)H_W(1-\gamma_W)]^{-1}$ is contained in
$H^2(\RR^3)$, the function $\rho_W$ belongs to $L^1(\RR^3) \cap H^2(\RR^3)$.
Using similar arguments, one can easily show that the operator $T_W$
defined by (\ref{eq:op_A}) is finite-rank and that $\rho_W(\br) =
t_W(\br,\br)$.  

\medskip

\noindent
It remains to prove Lemmas~\ref{lem:resolvent}
and~\ref{lem:traceclass}.

\medskip

\noindent
{\it Proof of Lemma~\ref{lem:resolvent}}.
Let $z$ be in the resolvent set $\rho(H_W)$ of $H_W$. By
Assumption~\ref{hyp:gap},
$D(H_W) = H^2(\RR^3)$. Hence, $(z-H_W)$, considered as an operator from
$H^2(\RR^3)$ to $L^2(\RR^3)$, is invertible. As $W \in L^2(\RR^3) +
L^\infty(\RR^3)$, it is also continuous, hence bicontinuous in view of
the inverse mapping theorem~\cite{ReedSimon}. As so is 
$(1-\Delta)$, $(1-\Delta)R^0(z)$ is a bounded operator on $L^2(\RR^3)$. 

\medskip

\noindent
On the other hand, it holds, for all $c > 0$ such that $(z-c) \in
\rho(H_W)$, 
$$
R^0(z)(1-\Delta) =
R^0(z)((z-c)-H_W)R^0(z-c)(c-\Delta/2)(c-\Delta/2)^{-1}(1-\Delta).
$$
The operators $R^0(z)((z-c)-H_W) = 1 - cR^0(z)$ and
$(c-\Delta/2)^{-1}(1-\Delta)$ are bounded operators on
$L^2(\RR^3)$. Besides, 
$$
(c-\Delta/2)^{-1}(z-c-H_W) = - \left(1- (c-\Delta/2)^{-1} (z-W) \right).
$$
As $W \in L^2(\RR^3)+L^\infty(\RR^3)$, one can write $W$ as
$W=W_2+W_\infty$ with $W_2 \in L^2(\RR^3)$ and $W_\infty \in
L^\infty(\RR^3)$. The operator $(c-\Delta/2)^{-1} (z-W_\infty)$ is
a bounded operator and its norm goes to zero when $c$ goes to
$+\infty$. Lastly, the operator $(c-\Delta/2)^{-1} W_2$ is
Hilbert-Schmidt, and its Hilbert-Schmidt norm
$$
\left\| (c-\Delta/2)^{-1} W_2 \right\|_\HS = \frac{1}{8\pi}
\left( \int_{\RR^3} \frac{{\rm e}^{-\sqrt{2c} \, |\br|}}{|\br|^2} \, d\br
\right)^{1/2} \|W_2\|_{L^2}, 
$$
hence its norm in ${\cal L}(L^2(\RR^3))$, go to zero when $c$ goes to
infinity. The operator $(c-\Delta/2)^{-1}(z-c-H_W)$ is therefore bounded
on $L^2(\RR^3)$ and invertible for $c$ large enough. Its inverse,
$R^0(z-c)(c-\Delta/2)$ also defines a bounded operator. This proves
that $R^0(z)(1-\Delta)$ is a bounded operator.

\medskip

\noindent
The analyticity of the functions $z \mapsto (1-\Delta)R^0(z)$ and 
$z \mapsto R^0(z)(1-\Delta)$ follows from the analyticity of the
resolvent on the resolvent set: For $z_0 \in \rho(H_W)$ and $z \in
\rho(H_W)$ such that $|z-z_0| < \|R^0(z_0)\|^{-1}$, it holds
$$
R^0(z) = \sum_{n=0}^{+\infty} (z-z_0)^n R^0(z_0)^{n+1}. 
$$
\cqfd

\medskip

\noindent
{\it Proof of Lemma~\ref{lem:traceclass}.} 
Let us first prove the second assertion. The kernel of the operator
$v(1-\Delta)^{-1}$ is explicit and reads
$$
k(\br,\br') = v(\br) \frac{{\rm e}^{-|\br-\br'|}}{4\pi|\br-\br'|}.
$$ 
As 
$$
\int_{\RR^3}\int_{\RR^3} k(\br,\br')^2 \, d\br \, d\br' = \left( \int_{\RR^3}
  v^2 \right) \, \left( \int_{\RR^3} \frac{{\rm e}^{-2|\br|}}{16\pi^2|\br|^2}
  \, d\br \right) = \frac{\|v\|_{L^2}^2}{8\pi},
$$
$v(1-\Delta)^{-1}$ is Hilbert-Schmidt and $\|v(1-\Delta)^{-1}\|_\HS =
\frac{\|v\|_{L^2}}{(8\pi)^{1/2}}$. 

\medskip

\noindent
In order to prove the second assertion, we write $v$ as $v=v_+-v_-$ with
$v_+ = \mbox{max}(v,0)$ and $v_-=\mbox{max}(-v,0)$, and introduce the
operators $A_{\pm} = \sqrt{v_{\pm}} (1-\Delta)^{-1}$. As $\sqrt{v_{\pm}} \in
L^2(\RR^3)$, the operators $A_{\pm}$ are Hilbert-Schmidt and such that
$\|A_{\pm}\|_\HS = \|\sqrt{v_{\pm}}\|_{L^2}/(8\pi)^{1/2}$. Hence,
$(1-\Delta)^{-1}v(1-\Delta)^{-1} = A_+^\ast A_+ - A_-^\ast A_-$ is
trace-class and
$$
\|(1-\Delta)^{-1}v(1-\Delta)^{-1}\| \le \frac{1}{8\pi} \left(
\|\sqrt{v_+}\|_{L^2}^2 + \|\sqrt{v_-}\|_{L^2}^2 \right) = 
\frac{\|v\|_{L^1}}{8\pi}. 
$$
\cqfd


\subsection{Proof of Proposition~\ref{prop:KLI}: Properties of the KLI
  potential} 

\noindent
As $D(H_W) = H^2(\RR^3)$, the eigenfunctions $\phi_i^W$ are in
$H^2(\RR^3)$, and are therefore continuous on~$\RR^3$.
Under the assumption that $\epsilon_1^W <
\mbox{min} \, \sigma_{\rm ess}(H_W)$, the ground state
$\phi_1^W$ is non-degenerate, and positive on $\RR^3$. Consequently,
$\rho_\Phi$ is continuous, and positive on $\RR^3$, so that 
\begin{equation} \label{eq:exp_KLI}
v_{x,{\rm KLI}}^{\Phi^W}(\br) = v_{x,S}^{\Phi^W}(\br) + \sum_{i=1}^{N} \left( 
\langle \phi_i^W \, | v_{x,{\rm KLI}}^{\Phi^W} \, | \, \phi_i^W \rangle
  - \langle \phi_i^W \, | K_{\Phi^W} \, | \, \phi_i^W \rangle \right) \,
\frac{|\phi_i^W(\br)|^2}{\rho_{\Phi^W}(\br)}, 
\end{equation}
is a continuous, bounded function on $\RR^3$. 

\medskip

\noindent
Proceeding as in Section~\ref{sec:proofs_Slater}, one can show that the
functional 
$J_\Phi^{\rm KLI}$ is well-defined on
$L^3(\RR^3)+L^\infty(\RR^3)$ and that the global minimizers $v$ of
(\ref{eq:var_KLI}) are exactly the solutions to the KLI equation 
\begin{equation} 
\label{eq:KLI_2}
\rho_{\Phi^W}(\br)  v(\br) = - \int_{\RR^3}
\frac{|\gamma_{\Phi^W}(\br,\br')|^2}{|\br-\br'|} \, d\br' + \sum_{i=1}^{N} 
\langle \phi_i^W \, | \, v - K_{\Phi^W} \, | \, \phi_i^W
\rangle  \, |\phi_i^W(\br)|^2.
\end{equation}
It remains to prove that the set of solutions to the above equation is a
one-dimensional affine space. To this end, we remark that a potential
$$
v(\br) = v_{x,S}^{\Phi^W}(\br) + \sum_{i=1}^{N} \left( \alpha^{\Phi^W}_i -
\langle \phi_i^W \, | \, K_{\Phi^W} \, | \, \phi_i^W
\rangle \right) \, \frac{|\phi_i^W(\br)|^2}{\rho_{\Phi^W}(\br)}
$$
is solution to (\ref{eq:KLI_2}) if and only is the vector $\alpha^{\Phi^W} =
(\alpha_i^{\Phi^W}) \in \RR^N$ is solution to the linear system
(\ref{eq:linear_system_KLI}). We 
therefore have to show that $\mbox{Ker}(I_N-S^{\Phi^W}) = \RR (1, \cdots , 1)^T$
and that $\beta^{\Phi^W} \in \mbox{Ran}(I_N-S^\Phi)$. 

\medskip

\noindent
Let $y \in \RR^N$. One has
$$
y^T (I_N-S^{\Phi^W}) y = \sum_{i=1}^N y_i^2 - 
\int_{\RR^3} \frac{\dps \left( \sum_{i=1}^N y_i (\phi_i^W)^2 \right)^2}
{\dps \sum_{i=1}^N (\phi_i^W)^2} \ge \sum_{i=1}^N y_i^2 - 
\int_{\RR^3} \frac{\dps \left( \sum_{i=1}^N y_i^2 (\phi_i^W)^2 \right) \, \left(
    \sum_{i=1}^N (\phi_i^W)^2 \right) }
{\dps \sum_{i=1}^N (\phi_i^W)^2} = 0,
$$
with equality if and only if, for all $\br \in \RR^3$, there exists
$\lambda(\br)$ such that $y_i \phi_i^W(\br) = \lambda(\br) \phi_i^W(\br)$ for all $1
\le i \le N$. As $\phi_1^W > 0$ on $\RR^3$, this condition is equivalent
to $y =(y_i) \in \RR (1, \cdots, 1)^T$. Thus, $\mbox{Ker}(I_N-S^{\Phi^W}) =
\RR (1, \cdots , 1)^T$. Lastly, using the fact that $S^{\Phi^W}$ is
symmetric, one obtains
$$
\mbox{Ran}(I_N-S^{\Phi^W}) = \mbox{Ker}(I_N-S^{\Phi^W})^\perp = \left\{ z =(z_i)
  \in \RR^N, \; \sum_{i=1}^N z_i=0 \right\}.
$$
It is easy to check that $\beta^{\Phi^W} \in \mbox{Ran}(I_N-S^{\Phi^W})$. 


\subsection{Proof of Proposition~\ref{prop:ELP}: Properties of the ELP}  

\noindent
For all $v \in L^3(\RR^3) + L^\infty(\RR^3)$, the operator $B^\Phi v =
[v,\gamma_\Phi]$ is Hilbert-Schmidt. One can therefore define on
$L^3(\RR^3) + L^\infty(\RR^3)$ the functional 
\[
J_\Phi^{\rm ELP}(v) = \frac12 \| [v-\KHF,\PHF] \|^2_\HS = 
\frac12 \| B^\Phi v - [\KHF,\PHF] \|^2_\HS.
\]
For all $v$ and $h$ in $L^3(\RR^3)+L^\infty(\RR^3)$, 
$$
J_\Phi^{\rm ELP}(v+h)  = J_\Phi^{\rm ELP}(v) +  \langle B^\Phi v -
[\KHF,\PHF] ,B^\Phi h \rangle_\HS + \frac 1 2 \| B^\Phi h \|^2_\HS,
$$
and
\begin{eqnarray*}
& &  \langle B^\Phi v - [\KHF,\PHF] ,B^\Phi h \rangle_\HS 
 \\ & = & 
2 \, \int_{\RR^3} 
\left( \rho_{\Phi}(\br)  v(\br) + \int_{\RR^3}
\frac{|\gamma_{\Phi}(\br,\br')|^2}{|\br-\br'|} \, d\br' - \sum_{i,j=1}^{N}
\langle \phi_i \, | v-K_\Phi  \, | \phi_j \rangle \phi_i(\br) \phi_j(\br)
 \right)  \, h(\br) \, d\br.
\end{eqnarray*}
The global minimizers $v$ of (\ref{eq:ELP_pb}) are therefore exactly the
solutions to the equation  
\begin{equation} 
  \label{eq:ELP_2}
  \rho_{\Phi}(\br)  v(\br) = - \int_{\RR^3}
  \frac{|\gamma_{\Phi}(\br,\br')|^2}{|\br-\br'|} \, d\br' + \sum_{i,j=1}^{N}
\langle \phi_i  | v-K_\Phi  | \phi_j \rangle \phi_i(\br) \phi_j(\br).
\end{equation}
Multiplying the above equation by $\frac{\phi_i\phi_j}{\rho}$ and
integrating over $\RR^3$, one then observes that a function $v$ satisfying 
$$
\rho_{\Phi}(\br)  v(\br) = - \int_{\RR^3}
\frac{|\gamma_{\Phi}(\br,\br')|^2}{|\br-\br'|} \, d\br' + \sum_{i,j=1}^{N}
(M_{ij} - \langle \phi_i | K_\Phi  | \phi_j \rangle) \phi_i(\br) \phi_j(\br)
$$
is solution to (\ref{eq:ELP_2}) if and only if the matrix $M$ is
solution to the linear system
\begin{equation}
\label{eq:Mv2}
(I-A^\Phi) M = G^\Phi.
\end{equation}
Let us now prove that, if the orbitals $\phi_i$ are continuous and if
$\RR^3 \setminus \rho_\Phi^{-1}(0)$ is connected, then $\mbox{Ker}(I -
A^\Phi) = \RR 
I_N$ and $G^\Phi \in \mbox{Ran}(I - A^\Phi)$. For this purpose,
let us consider a matrix $M \in {\mathcal M}_S(N)$ such that
$\left( I - A^\Phi \right) M = 0$.
As $M$ is symmetric, it can be diagonalized in an orthonormal basis
set as
\[
M = U^T \; \mbox{Diag}(\lambda_1, \cdots, \lambda_N) \; U
\]
where $U$ is a unitary matrix. Denoting by $(\psi_1,\dots,\psi_N)^T = U
(\phi_1,\dots,\phi_N)^T$, a simple calculation leads to
\[
0 = ( \left( I - A^\Phi \right) M, M)_F =
\sum_{i=1}^N \lambda_i^2 -
\dps \int_\mR \left|  \sum_{i=1}^N \lambda_i 
    \psiHF_i(\br)^2 \right|^2 \, \frac{d\br}{\rhoHF(\br)},
\]
where $(\cdot, \cdot)_F$ is the Frobenius inner product on 
${\mathcal M}_S(N)$. As $U$ is a unitary transform, 
the $\psiHF_i$ are orthonormal for the
$\LL^2(\mR)$ inner product and $\displaystyle \sum_{i=1}^N \psiHF_i(\br)^2 =
\rhoHF(\br)$. Therefore, using Cauchy-Schwarz inequality, 
\[
\left| \sum_{i=1}^N \lambda_i  \psiHF_i(\br)^2 \right|^2 
\le \left( \sum_{i=1}^N \psiHF_i(\br)^2 \right)  \; \left( \sum_{i=1}^N
\lambda_i^2 \psiHF_i(\br)^2 \right) = \rhoHF(\br) \, 
\sum_{i=1}^N \lambda_i^2 \psiHF_i(\br)^2, 
\]
with equality if and only if there exists $C(\br)$
such that $\lambda_i \psiHF_i(\br) = C(\br)  \psiHF_i(\br)$ for all $1 \le i
\le N$.  
Hence,
\[
\sum_{i=1}^N \lambda_i^2 -
\int_\mR \displaystyle \left|  \sum_{i=1}^N \lambda_i 
    \psiHF_i(\br)^2 \right|^2 \, \frac{d\br}{\rhoHF(\br)} \ge 
\sum_{i=1}^N \lambda_i^2 - \int_\mR \sum_{i=1}^N \lambda_i^2 \psiHF_i^2
 = 0,
\]
with equality if and only if for almost all $\br \in \mR$, there exists  $C(\br)$
such that $\lambda_i \psiHF_i(\br) = C(\br)  \psiHF_i(\br)$ for all $1 \le i
\le N$. 

\medskip

\noindent
If the orbitals $\phi_i$ are continuous, so are the functions
$\psi_i$. Let us consider the open sets $\Omega_i = \RR^3 \setminus
\psi_i^{-1}(0)$ and $\Omega = \cup_{i=1}^N \Omega_i = \RR^3 \setminus
\rho_\Phi^{-1}(0)$. On $\Omega_i$, one has $C(\br) =
\lambda_i$. This implies that the function $C(\br)$ is constant on each connected
component of $\Omega$. If $\Omega$ is connected, one therefore has
$\lambda_1 = \lambda_2  = \cdots = \lambda_N$, i.e.  
$M$ is proportional to the identity matrix. 

\medskip

In summary, under the assumptions that the orbitals $\phi_i$ are
continuous and that $\RR^3 \setminus \rho_\Phi^{-1}(0)$ is connected,
\begin{enumerate}[\quad (1)]
\item the linear equation (\ref{eq:Mv2}) has a solution if and only if 
$G^\Phi \in \mbox{Ran} \left( I - A^\Phi \right)$.
Note that $\mbox{Ran} \left( I - A^\Phi \right) = 
\mbox{Ker} \left( I - (A^\Phi)^\ast \right)^\perp = 
\mbox{Ker} \left( I - A^\Phi \right)^\perp$,
since $A^\Phi$ is self-adjoint for the Frobenius inner product. It then
follows that $\mbox{Ran} \left( 
I - A^\Phi \right) = {\rm
  Span}(I_{N})^\perp$. Since $(I_N, G^\Phi)_F = {\rm Tr}(G^\Phi) = 0$,
$G^\Phi \in \mbox{Ran}
\left( I - A^\Phi \right)$ and (\ref{eq:Mv2}) has at
least one solution $M^\Phi_\star$;
\item if $M^\Phi_\star$ is a solution to (\ref{eq:Mv2}), then the set of the
  solutions of (\ref{eq:Mv2}) is $\left\{M^\Phi_\star + \lambda I_{\RR^N}, \;
  \lambda \in \RR \right\}$.
\end{enumerate}
Note that replacing $M^\Phi$ with $M^\Phi+\lambda I_{\RR^N}$
in (\ref{eq:Mv2}) amounts to replacing $v^\Phi_\rx$ with $v^\Phi_\rx +
\lambda$. 

\section*{Appendix: Elements of functional analysis}

\noindent
This Appendix summarizes the basic concepts of functional analysis needed
to understand the {\em statements} of the results contained in the
present article. The additional concepts and results used in the proofs
can be found in Ref.~\cite{ReedSimon}.

\medskip

\noindent
Let us first recall the definition of the functional spaces used
throughout this article. In the following, all the considered functions
are real-valued Lebesgue measurable functions on $\RR^3$. As usual, two
functions which differ only on a set of measure zero are identified.

\medskip

For $1 \le p < \infty$, the $L^p$ space is defined as
$$
L^p(\RR^3) = \left\{ u \; \left | \; \int_{\RR^3} |u(\br)|^p \, d\br 
< \infty \right. \right\}. 
$$
Endowed with the norm 
$$
\| u \|_{L^p} = \left( \int_{\RR^3} |u(\br)|^p \, d\br \right)^{1/p}, 
$$
$L^p(\RR^3)$ is a Banach space. The space 
$L^2(\RR^3)$ is a Hilbert space for the inner product
$$
\langle u | v \rangle = \int_{\RR^3} u(\br) \, v(\br) \, d\br.
$$
The space $L^\infty(\RR^3)$ is the
vector space of essentially bounded functions. A measurable function $u$
is essentially bounded if there exists a constant $M$ such that $|u| \le
M$ almost everywhere (a.e.), i.e. everywhere except, possibly, on a set
of measure zero. Endowed with the norm 
$$
\| u \|_{L^\infty} = \inf \left\{ M \ge 0 \; | \; |u| \le M \mbox{ a.e.}
\right\},
$$
$L^\infty(\RR^3)$ is a Banach space. One has for all $u \in
L^\infty(\RR^3)$, 
$$
|u(\br)| \le \|u\|_{L^\infty} \quad \mbox{a.e.}
$$
For all $1 \le p < q \le \infty$, the space $L^p(\RR^3)\cap L^q(\RR^3)$,
endowed with the norm $\|\cdot\|_{L^p\cap L^q} =
\|\cdot\|_{L^p}+\|\cdot\|_{L^q}$, is a Banach space. Likewise, for all
$1 < p < q \le \infty$ the space
$$
L^p(\RR^3)+L^q(\RR^3) =
\left\{u \; | \; \exists (u_p,u_q) \in L^p(\RR^3) \times L^q(\RR^3), \;
  u = u_p+u_q \right\},
$$
endowed with the norm
$$
\|u\|_{L^p+L^q} = \inf \left\{ \|u_p\|_{L^p}+\|u_q\|_{L^q}, \; 
(u_p,u_q) \in L^p(\RR^3) \times L^q(\RR^3), \;
  u = u_p+u_q \right\}
$$
is a Banach space.

\medskip

In quantum mechanics, the kinetic energy of a one-particle wavefunction
$\phi$ is $\frac 1 2 \int_{\RR^3} |\nabla \phi|^2$. It is therefore
natural to introduce the vector space
$$
H^1(\RR^3) = \left\{ u \in L^2(\RR^3) \; | \; \nabla u \in
  (L^2(\RR^3))^3 \right\}.
$$
Endowed with the inner product
$$
(u,v)_{H^1} = \int_{\RR^3} u(\br) \, v(\br) \, d\br  + 
\int_{\RR^3} \nabla u(\br) \cdot \nabla v(\br) \, d\br,
$$
$H^1(\RR^3)$ is a Hilbert space. We will also use the Hilbert space 
$$
H^2(\RR^3) = \left\{ u \in H^1(\RR^3) \; \left | \, \forall \, 1 \leq i,j
\leq 3, \ \frac{\partial^2 u}{\partial
    \br_i \partial \br_j} \in L^2(\RR^3) \right. \right\}  
$$
whose inner product is 
$$
(u,v)_{H^2} = \int_{\RR^3} u(\br) \, v(\br) \, d\br + \int_{\RR^3}
\nabla u(\br) \cdot \nabla v(\br) \, d\br + 
\sum_{i,j=1}^3 \int_{\RR^3} \frac{\partial^2 u}{\partial \br_i \partial
  \br_j}(\br) \frac{\partial^2 v}{\partial \br_i \partial \br_j}(\br) \,
d\br.
$$
The functional spaces $H^1(\RR^3)$ and $H^2(\RR^3)$ belong to the class
of Sobolev spaces.  

\medskip

Lastly, $L^p_{\rm loc}(\RR^3)$ is the vector space of the
functions $u$ such that $\int_K |u(\br)|^p \, d\br < \infty$ for all compact sets $K
\subset \RR^3$.

\medskip


The second part of this Appendix is devoted to linear operators on $L^2(\RR^3)$.
The set of the continuous linear operators from $L^2(\RR^3)$ to
$L^2(\RR^3)$, also called bounded operators on $L^2(\RR^3)$, 
is denoted by ${\cal L}(L^2(\RR^3))$. The adjoint of a
continuous linear operator $T \in {\cal L}(L^2(\RR^3))$ is the unique
operator of ${\cal L}(L^2(\RR^3))$, denoted by $T^\ast$, defined by 
$$
\forall (u,v) \in L^2(\RR^3) \times L^2(\RR^3), \quad \langle T^\ast u |
v \rangle = \langle u | Tv \rangle. 
$$
The operator $T \in {\cal L}(L^2(\RR^3))$ is called self-adjoint if $T^\ast=T$.
The vector space of self-adjoint continuous linear operators on
$L^2(\RR^3)$ is denoted by ${\cal S}(L^2(\RR^3))$. If $T$ is a
self-adjoint operator, it is usual to denote by
$$
\langle u|T|v \rangle = \langle Tu|v\rangle = \langle u |Tv \rangle.
$$

Let $T \in {\cal L}(L^2(\RR^3))$ and $(e_n)_{n \in \NN}$ a Hilbert basis
of $L^2(\RR^3)$. The value of the sum
$$
\sum_{n \in \NN} \|Te_n\|_{L^2}^2
$$
is independent of the choice of the Hilbert basis $(e_n)_{n \in \NN}$.
The operator $T$ is called Hilbert-Schmidt if
$$
\|T\|_\HS := \left(\sum_{n \in \NN} \|Te_n\|_{L^2}^2\right)^2 < \infty.
$$
The set of Hilbert-Schmidt operators forms a vector space, denoted by
$\HS$. It is in fact a Hilbert space for the inner product
$$
\langle S,T \rangle_\HS = \sum_{n \in \NN} \langle Se_n|Te_n \rangle.
$$
The norm associated with $\langle \cdot,\cdot
\rangle_\HS$ is denoted by $\| \cdot \|_\HS$.
It can be proved that $T \in {\cal L}(L^2(\RR^3))$ is Hilbert-Schmidt if 
and only if there exists a function of $L^2(\RR^3 \times \RR^3)$, called 
the kernel of the operator $T$ and usually denoted by $T$ as well, such
that 
$$ 
(Tu)(\br) = \int_{\RR^3} T(\br,\br') \, u(\br') \, d\br'. 
$$ 
It holds
$$ 
\| T \|_\HS = \left( \int_{\RR^3 \times \RR^3} |T(\br,\br')|^2 \, d\br \, d\br' 
\right)^{1/2}, 
$$ 
and $T$ is self-adjoint if and only if $T(\br',\br) = T(\br,\br')$.

Let now $T \in {\cal S}(L^2(\RR^3))$ be a non-negative self-adjoint operator 
(i.e. $\langle u|T|u \rangle \ge 0$ for all $u \in L^2(\RR^3)$) and $(e_n)_{n \in \NN}$ 
a Hilbert basis of $L^2(\RR^3)$. The value of the sum 
$$ 
\sum_{n \in \NN} \langle e_n|T|e_n \rangle 
$$  
does not depend on the choice of the Hilbert basis $(e_n)_{n \in 
  \NN}$. If this sum is finite, $T$ is called trace-class and the trace 
of $T$ is defined as 
$$ 
\Tr (T) = \sum_{n \in \NN} \langle e_n|T|e_n \rangle. 
$$ 
A (non-necessarily self-adjoint) operator $T \in {\cal L}(L^2(\RR^3))$ is 
called trace-class if the non-negative self-adjoint continuous operator 
$|T|=(T^\ast T)^{1/2}$ is trace-class (the square root of a non-negative
self-adjoint operator is defined below). The set of trace-class
operators on $L^2(\RR^3)$ forms a vector subspace of $\HS$, denoted by
$\tc$. Endowed with the norm 
$$
\|T\|_{\tc} = \Tr(|T|),
$$
$\tc$ is a Banach space. For all $T \in \tc$, the sum 
$\sum_{n \in \NN} \langle e_n |Te_n\rangle$ is finite and independent of
the choice of 
the Hilbert basis $(e_n)_{n \in \NN}$. The trace $\Tr$ defines a
continuous linear form on $\tc$.

\medskip

Most linear operators arising in quantum mechanics are not continuous
linear operators. An example is the one-particle kinetic energy operator
$T_K = -\frac 1 2 \Delta$. As the Laplacien of a function of
$L^2(\RR^3)$ is not, in general, a function of $L^2(\RR^3)$, $T_K$
cannot be defined 
as a linear application from $L^2(\RR^3)$ to itself. The useful
definition of linear operators is the following: A linear operator $T$ on
$L^2(\RR^3)$ is a $L^2(\RR^3)$-valued linear application defined on a
subspace $D(T)$ of $L^2(\RR^3)$. The set $D(T)$ is called the domain of the
linear operator $T$. For instance $T_K$ is a linear operator on
$L^2(\RR^3)$ with domain $D(T_K) = H^2(\RR^3)$ (for all $u \in
H^2(\RR^3)$, $\Delta u \in L^2(\RR^3)$, and $T_Ku$ therefore is a function
of $L^2(\RR^3)$).

Let $T$ be a linear operator on $L^2(\RR^3)$ with dense domain
$D(T)$. The adjoint of $T$ is the unique linear operator on $L^2(\RR^3)$
defined by
\begin{eqnarray*}
D(T^\ast) & \!\!\! = \!\!\! & \left\{ u \in L^2(\RR^3) \; | \; \exists v_u \in L^2(\RR^3)
  \mbox{ such that } \langle v_u|w \rangle = \langle u|Tw \rangle \;
  \forall w \in D(T) 
\right\} \\ T^\ast u & \!\!\! = \!\!\! & v_u \qquad \mbox{($v_u$ is
  uniquely defined 
  since $D(T)$ is dense in $L^2(\RR^3)$)}.
\end{eqnarray*}
The operator $T$ is called self-adjoint if $T^\ast = T$ (i.e. if $D(T^\ast)=D(T)$ and
if for all $u \in D(T) = D(T^\ast)$, $T^\ast u = Tu$). 

\medskip

\noindent
Let $T$ be a self-adjoint operator on $L^2(\RR^3)$ with domain $D(T)$,
and $z \in \CC$. In order to simplify the notation, we denote by $z -
T$ the operator $zI_{L^2(\RR^3)} - T$ where $I_{L^2(\RR^3)}$ is the
identity operator on $L^2(\RR^3)$. If $z-T$ is an invertible
operator from $D(T)$ to $L^2(\RR^3)$, it can be proved that
$R(z)=(z-T)^{-1}$ defines a continuous linear operator on $L^2(\RR^3)$
(with range $D(T)$). The set
$$
\rho(T) = \left\{ z \in \CC \; | \; \mbox{$z-T$ is an invertible
operator from $D(T)$ to $L^2(\RR^3)$} \right\}
$$ 
is called the resolvent set of $T$, and the family $(R(z))_{z \in
  \rho(T)}$ the resolvent of $T$. The spectrum of $T$ is the set
$\sigma(T) = \CC \setminus \rho(T)$. The set $\rho(T)$ is an open set of $\CC$
and $\sigma(T)$ is a closed subset of $\RR$. An eigenvalue of $T$ is a
complex number $\lambda$ for which there exists $u \in L^2(\RR^3)$ such
that $Tu=\lambda u$. The set of all the eigenvalues of $T$ is called the
point spectrum of $T$ and is denoted by $\sigma_{\rm p}(T)$. Obviously,
$\sigma_{\rm p}(T) \subset \sigma(T)$ (in particular, all the
eigenvalues of a self-adjoint operator are real). The set $\sigma_{\rm
  c}(T) = \sigma(T) \setminus \sigma_{\rm p}(T)$ is called the continuous
spectrum of $T$. If the continuous spectrum of $T$ is empty (i.e. if 
$\sigma(T) = \sigma_{\rm p}(T)$), there exists a Hilbert basis
$(e_n)_{n \in \NN^\ast}$ of $L^2(\RR^3)$ which diagonalizes~$T$:
$$
T = \sum_{n \in \NN} \lambda_n \langle e_n | \, \cdot \, \rangle e_n \qquad \left( =
  \sum_{n \in \NN} \lambda_n \, |e_n \rangle \, \langle e_n| \quad \mbox{ in
    bra-ket notation} \right),
$$
with $\lambda_n \in \RR$. In this case $\sigma(T) = \left\{ \lambda_n
\right\}$. If $f \, : \, \RR \longrightarrow \CC$ is continuous in a
neighborhood of $\sigma(T)$, the operator $f(T)$ is defined as
\begin{eqnarray*}
D(f(T)) & \!\!\! = \!\!\! & \left\{ u = \sum_{n \in \NN} u_n e_n \in
  L^2(\RR^3) \; \left | \; \sum_{n \in \NN} (1+|f(\lambda_n)|^2) |u_n|^2 < \infty
\right. \right\} \\ 
f(T)  & \!\!\! = \!\!\! & \sum_{n \in \NN} f(\lambda_n) 
\langle e_n | \cdot \rangle e_n =  \sum_{n \in \NN} f(\lambda_n) \, |e_n \rangle
\, \langle e_n|. 
\end{eqnarray*}
This definition can be generalized to any self-adjoint operator $T$
by means of the Spectral Theorem~\cite{ReedSimon}. Note that if $T$ is
non-negative, $\sigma(T) \subset \RR_+$ and the operator $T^{1/2}$ can
therefore be given a sense.

Lastly, the spectrum $\sigma(T)$ of a self-adjoint operator can also be
partitioned as follows
$$
\sigma(T) = \sigma_{\rm d}(T) \cup  \sigma_{\rm ess}(T),
$$
where $\sigma_{\rm d}(T)$ is the set of all the isolated eigenvalues of
$T$ of finite multiplicity, and where $\sigma_{\rm ess}(T) = \sigma(T)
\setminus \sigma_{\rm d}(T)$. The sets $\sigma_{\rm d}(T)$ and $\sigma_{\rm
  ess}(T)$ are called respectively the discrete spectrum and the
essential spectrum of $T$.

\section*{Acknowledgments} We warmly thank C. Le Bris and M. Lewin for
useful discussions and comments.

\bibliographystyle{gabi}
\bibliography{exchange.bib}

\end{document}